\RequirePackage{fix-cm}
\documentclass[smallcondensed]{svjour3}     % onecolumn (ditto)
\smartqed  % flush right qed marks, e.g. at end of proof
\usepackage{graphicx}
 \usepackage{mathptmx}      % use Times fonts if available on your TeX system
\usepackage{amssymb}
\usepackage{amsmath}
\usepackage{natbib}
\usepackage{hyperref}
\usepackage{longtable}
\usepackage{xspace}
\newcommand{\sign}{\mathop{\mathrm{sign}}\nolimits}
\newcommand{\FAP}{\mathrm{FAP}}

 \journalname{Celestial Mechanics \& Dynamical Astronomy}
\begin{document}

\title{Analysing the Main Belt asteroid distributions by wavelets
}


\author{R.V.~Baluev$^\star$ \and E.I.~Rodionov}

\authorrunning{Baluev \& Rodionov} % if too long for running head

\institute{R.V.~Baluev$^\star$ \and E.I.~Rodionov \at
              Saint Petersburg State University, Universitetskaya emb. 7--9, St Petersburg 199034, Russia \\
              \email{r.baluev@spbu.ru} (R.V.~Baluev)
}

\date{Received: 08.01.2020 / Accepted: date}

\maketitle

\begin{abstract}
We perform statistical wavelet analysis of the Main Belt asteroids, seeking statistically
significant asteroid families. The goal is to test the new wavelet analysis algorithm and
to compare its results with more traditional methods like the hierarchic clustering. We
first consider several 1D distributions for various physical and orbital parameters of
asteroids. Then we consider three bivariate distributions for the three orbital parameters
$(a,e,i)$ taken pairwisely. The full 3D analysis of this space is not available here, but
based on the 2D results we perform a disentangling of overlapped 2D families and derive
total of $44$ 3D families with confirmed statistical significance.
\end{abstract}

\section{Introduction}
\label{sec_intro}
First attempts to find asteroid families date back to XIX century, although the discovery
of new asteroids was rather slow that time, compared to the contemporary rate. In 1876,
based on just about 150 asteroids, D.~Kirkwood noticed about 10 asteroid groups, each
containing just 2-3 members moving along similar orbits. It was suggested that asteroids in
such groups may have a common origin, e.g. are fragments of larger disrupted bodies. F.
Tisserand continued Kirkwood attempts, composing a list of 417 asteroids (1891), and also
introducing a formal orbital classification characteristic now well-known as the Tisserand
invariant $T_J$ \citep[see][]{Hirayama22}. The number of asteroid families grew as new
asteroids were discovered. However, no other factors supported the common origin within a
family, except for orbit closeness. Therefore, the physical relationship of such asteroids
often remained too disputable.

Later on, \citet{Hirayama18} noticed that it might be not reasonable to compare
\emph{contemporary} orbits of asteroids for that goal. On a long time scale, planetary
perturbations may change orbits a lot, even if asteroids indeed were fragments of the same
body in some past and had close orbits initially. This motivation leaded K.~Hirayama to the
idea of invariant orbital elements that would remain (nearly) constant regardless of the
planetary perturbations. Hirayama constructed such invariant elements based on the Lagrange
perturbation theory and introduced them as `proper elements', assuming that asteroids from
the same family inherited them from their progenitors, should the latter existed in some
past. The proper elements remain very useful to identify the asteroid families.

Of course, the fragments may attain some minor additional velocities after a disruption of
a larger body, still resulting in some minor spread even in terms of the proper elements.
Moreover, the boundaries of such families are typically rather vague, merging with the
background distribution. Because of this, the population even within well-known asteroid
families is difficult to determine strictly, and it is often uncertain whether an asteroid
belongs to a particular family or not.

Although Hirayama introduced the notion of `asteroid family' in the sense of an asteroid
group probably sharing common origin, nowadays this term is relatively ambiguous. Even the
closeness of proper elements of some asteroids does not guarantee their common origin.
Other explanations are also possible, for example the mean-motion resonance (MMR). In this
case an MMR may serve as an orbital `trap' capturing objects that would not otherwise have
any common history. Possible example might be e.g. the Hilda family \citep{Broz08}, though
presently it is deemed to be a widely eroded superfamily of common origin \citep{Milani17}.
In some part, the question of common origin of asteroids may be resolved using their
spectral classification, but given the current high discovery rate it is not feasible to
perform an accurate taxonomic analysis of all asteroids \citep{Masiero15}. Also, the
disrupted progenitor body could be so large that its fragments would appear chemically
different.

Therefore, in this work under a `family of asteroids' we understand a group of objects
simply having close orbital or physical parameters. Such a property may \emph{suggest a
hint} that these asteroids could have common evolutionary origin, but does not guarantee
that.

Presently, the most popular method of asteroid family identification is the hierarchic
clustering method (HCM hereafter), which looks for objects with a small distance between
each other or from a main asteroid \citep{Zappala90}. The advantage of this method is that
it does not explicitly specifies any assumptions about the shape of the asteroid family in
the space of orbital parameters. Also, it can be relatively easily extended to higher
dimensions (larger number of the parameters involved). Its main disadvantage, which becomes
increasingly important when more and more small asteroids are discovered, is the effect of
`chaining'. In case of a collision, small fragments are thrown away with larger velocities,
and also they are subject to a stronger Yarkovsky effect. Therefore, they spread further in
the parametric space, revealing a tendency to distribute more uniformly and create
`bridges' between different families. This issue is currently solved using rather
artificial methods, e.g. by cutting the parametric space into distinct domains. Finally,
the results may differ depending on the orbital distance metric \citep{Nesvorny15}.

Yet another method of asteroid family identification is wavelet analysis, which was used
previously but did not attain the same popularity so far. Based on a sample of $\sim 12000$
asteroids, \citet{Zappala95} concluded that HCM and wavelet analysis methods yield similar
results. Both methods detected the same families, though the number of asteroids in a
family was different. We believe that the wavelet analysis was abandoned after that
because, firstly, it was a pretty young technique at that time (especially in what concerns
statistical tasks), and secondly, because it is more computationally demanding.

But now the wavelet analysis gained a considerably wider attention, whereas the computing
hardware progressed greatly. Also, the mathematical theory of the \emph{statistical}
flavours of the wavelet analysis was significantly improved, compared to 1990s. For
example, in \citep{Baluev18a} a new algorithm was presented, targeting the analysis of 1D
statistical distributions. It is not a cluster detection algorithm in the common sense,
because it has a more wide range of applications than just clusters identification. In
particular, it allows to detect distribution gaps as well as clusters, and also to
investigate the finely-resolved distribution shape inside the cluster (or gap). Contrary to
methods from \citep{Zappala95}, the primary attention is paid to the optimised statistical
sensitivity to allow a detection of patterns with smaller S/N ratio. Moreover, the
significance of the detected patterns is expressed in the traditional and intuitive
`p-value' or `n-sigma' notation. In fact, the algorithm from \citep{Baluev18a} represents a
tool to clean the shot noise (or finite sample noise) from an estimated density function,
based on certain predefined statistical tolerance, and aiming to detect patterns of certain
shape determined by the selected wavelet.

This technique is under a continuous development, for example the 2D analysis tool was
constructed recently \citep{BalRodShai19}, and further generalisations are also possible.
In this work, we aim to further develop this wavelet analysis method and the associated
software, presenting our analysis results regarding the numerous sample of the Main Belt
asteroids.

Notice that we do not advocate that wavelet analysis may (or is expected to) supersede the
HCM in some concern. Rather, they represent two qualitatively different methods of the
analysis. Their detailed comparison in terms of reliability and efficiency is definitely
interesting, but this is too big task for this paper. Instead, we only plan to provide some
field testing of our wavelet algorithm in the asteroid analysis task.

Very recently, machine learning was also introduced for asteroid family detection
\citep{Carruba19}, though in this work we omit detailed discussion of methods of this type.

In Sect.~\ref{sec_pa} we discuss some details about proper elements and the asteroid
families. In Sect.~\ref{sec_samples} our asteroid samples are discussed. In
Sect.~\ref{sec_statwa} we give several basic details about our wavelet analysis algorithm.
In Sect.~\ref{sec_1D} we analyse 1D distributions of various asteroid parameters. In
Sect.~\ref{sec_2D} we present results of the 2D analysis and a list of detected asteroid
families in the 3D space of semimajor axis, eccentricity, and inclination.

\section{Scientific context regarding the proper elements}
\label{sec_pa}
The classic definition says that proper orbital elements are quasi-integrals of the motion
equations, so they remain almost constant in time \citep{Knezevic02}. Proper elements can
be obtained after removal of short- and long-period perturbations from their osculating
counterparts and hence represent some ``mean'' motion characteristics. \citep{Hirayama18},
who introduced the concept of proper elements, also showed that some asteroids tend to
accumulate into groups in the plains $(a,e_p)$ and $(a,i_p)$, where $e_p$ and $i_p$ are
proper eccentricity and proper inclination. He supposed that such groups, or
\textit{families}, formed as a result of disintegration of a large parent body.
Each family was named based on its largest member object.

Hirayama introduced in his works the notions of a proper eccentricity and proper
inclination, but not of the proper semimajor axis, because the latter one has no secular
perturbations \citep{MurrayDermott}. Nevertheless, the contemporary notion of proper
semimajor axis includes averaging with respect to short-period perturbations
\citep{Knezevic02}.

In the classic theory of perturbations, the eccentric variables $h = e\sin{\varpi}$ and $k
= e\cos{\varpi}$ vary along a nearly-circular curve in the 2D plane, with a constant
angular velocity. Then, the proper eccentricity has an easy interpretation. The center of
the circle (the so-called forced eccentricity) would move along a complicated trajectory
defined by secular perturbations (and depending on semimajor axes), while the radius of the
circle is equal to the proper eccentricity, which is defined by initial conditions and thus
represents a fundamental orbital parameter. An analogous interpretation can be used for the
inclinational parameters $\sin i$ and $\Omega$ \citep[chap.~7]{MurrayDermott}.

Two methods are currently used to determine the proper elements: the analytic and so-called
synthetic ones \citep{Knezevic02}. The first method is based on the perturbation theory and
involves the computation of averaged elements using canonical transform to remove short-
and long-periodic terms. Synthetic theory involves the integration of asteroid and
planetary motion and filtering of periodic perturbations. After that the primary harmonics
are determined by means of the Fourier analysis. These primary harmonics are the proper
elements.

The synthetic method is now more suitable thanks to its better accuracy \citep{Knezevic02}.
Additionally, the analytic removal of short- and long-period term becomes difficult near
the resonances, so the synthetic method performs better in such conditions. The resonant
proper elements may be obtained by means of resonance averaging.

\begin{figure}
\includegraphics[angle=270,origin=c,width=\textwidth]{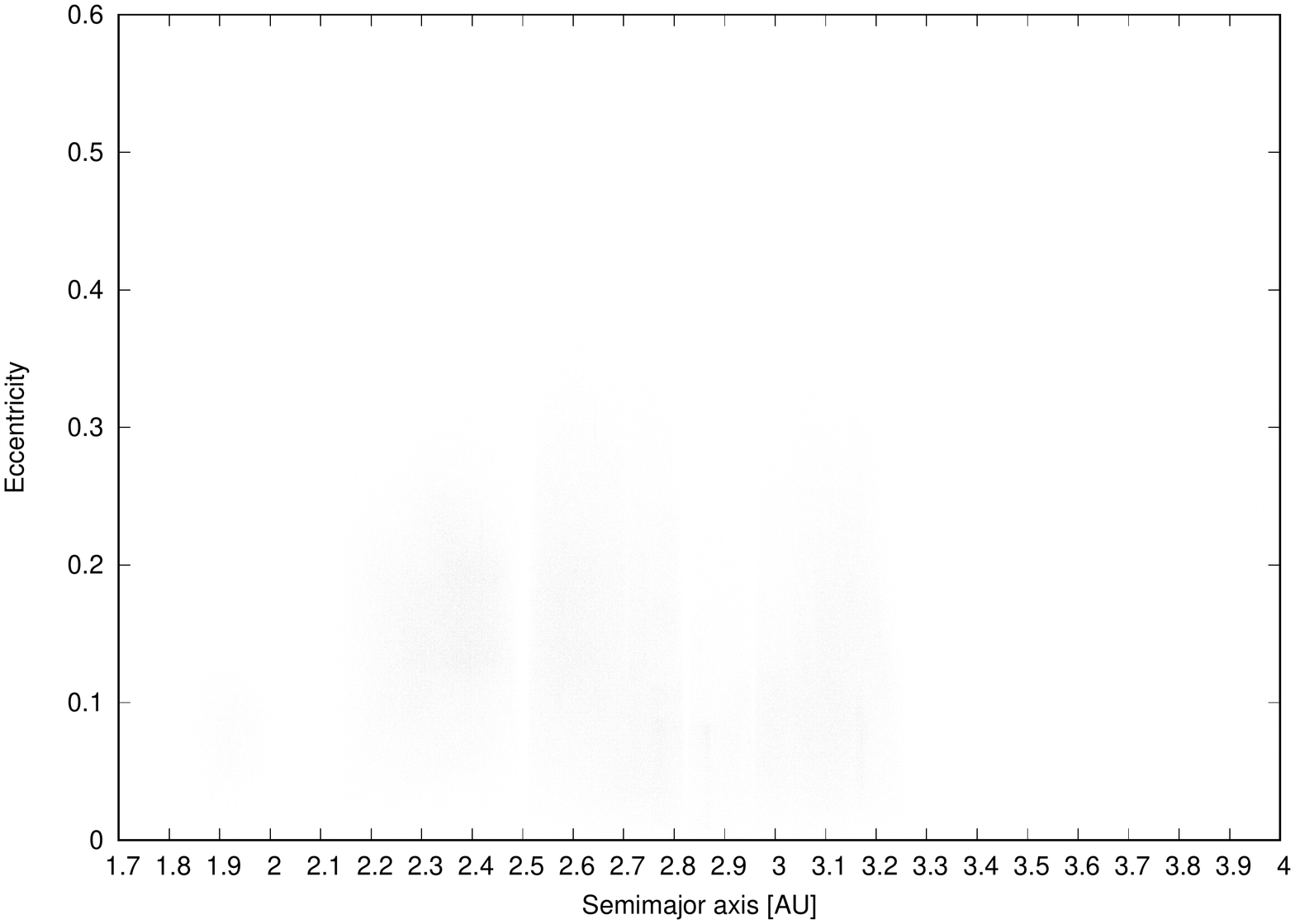}
\caption{Diagram for semimajor axis -- eccentricity (osculating) for the Main Belt. We can
see vertical bands marking the mean-motion resonances.}
\label{All_Osc}
\end{figure}

\begin{figure}
\includegraphics[angle=270,origin=c,width=\textwidth]{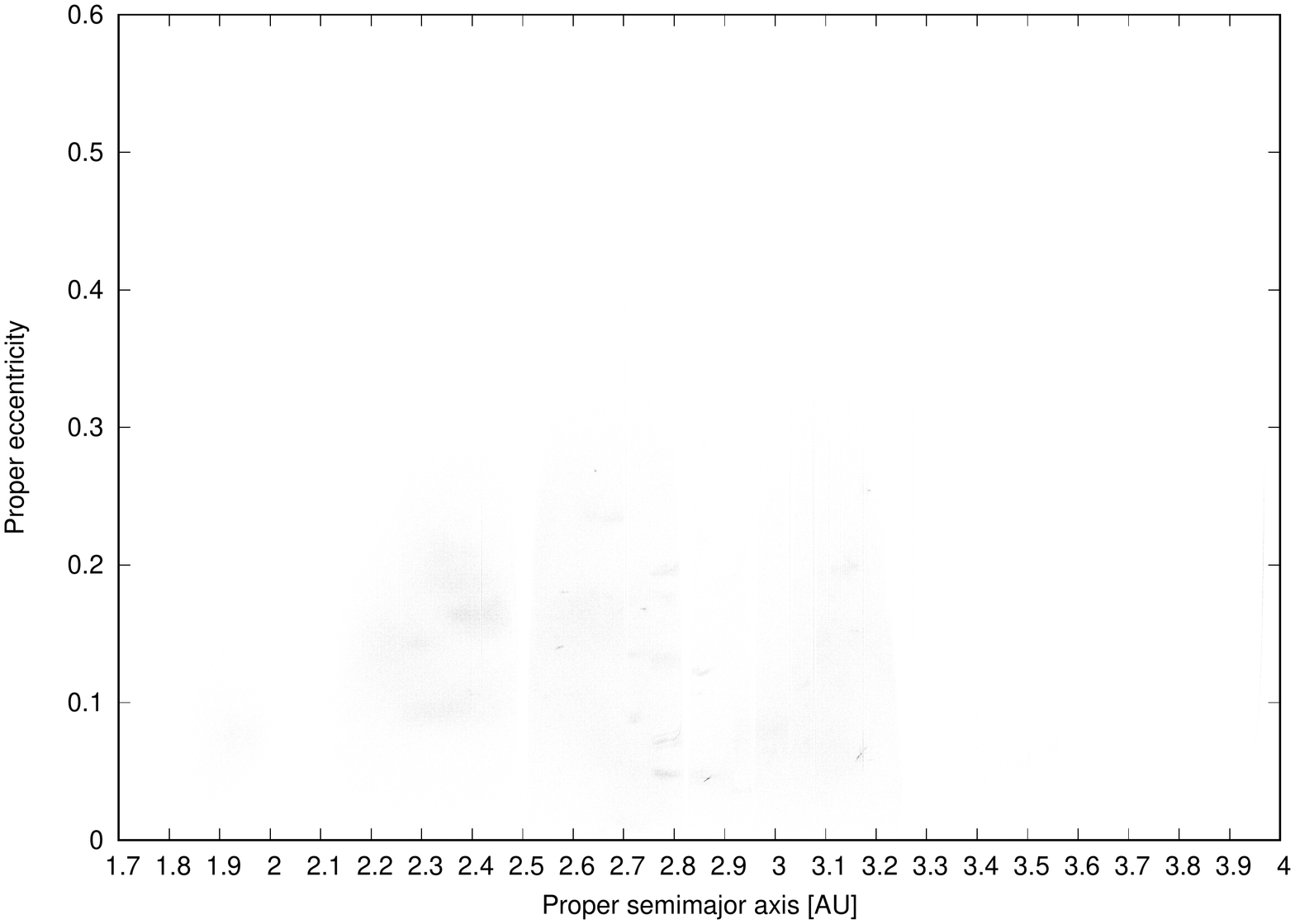}
\caption{Diagram for proper semimajor axis -- proper eccentricity for the Main Belt.
Contrary to Fig.~\ref{All_Osc}, we can clearly see compact spots additionally to vertical
bands.}
\label{All_Syn}
\end{figure}

In Fig. \ref{All_Syn} that shows a 2D distribution of asteroids in the $a$--$e$ plane, we
can see multiple vertical bands, which mark various mean-motion resonances that dominate in
the dynamical regime of the relevant asteroids. We notice that our 1D wavelet analysis
algorithm \citep{Baluev18a} suits quite well to locate such 1D bands.

Unfortunately, neither analytic nor synthetic method can process the secular resonances
well, e.g. the cases when two orbits have synchronized precession rates (for their
perihelia, or ascending nodes, or both). Such asteroids have less accurate estimations of
proper elements (especially proper eccentricity). In this work we removed from our analysis
the asteroids in the secular resonance $g+g_5-2g_6$, which are explicitly identified in
AstDys.

As we can see from Fig.~\ref{All_Syn}, the proper elements allow to identify many asteroid
families even by a plain look. However, some more subtle families may be more hard to
reveal in the background distribution, emphasizing the value of a sophisticated statistical
analysis in this task.

\section{Asteroid samples}
\label{sec_samples}
Though there were attempts to reveal families among the trans-Neptune objects
\citep{Snogdrass12}, such objects are relatively few, and their orbital elements are less
accurate, so in this work we limited ourselves mostly to the Main Belt.

The osculating orbital elements and physical parameters were taken from the Lowell
observatory catalog \texttt{astorb.dat} as of February 2019
\footnote{\texttt{http://asteroid.lowell.edu/main/astorb}}. For our analysis we used only
numbered asteroids in the range from $0.7$~AU to $7$~AU in semimajor axis. In these limits
we had $523175$ objects with known eccentricity, inclination, and absolute magnitude.
Unfortunately, the color index B-V was known only for $941$ asteroids, while the diameter
was known for $2139$ ones. Nonetheless, we tried to analyse the distributions for all these
physical parameters too, since based on the previous experience \citep{Baluev18b}, samples
containing 1-2 thousand of objects still may reveal statistical clusters.

The proper elements are available in the AstDys database
\footnote{\texttt{http://newton.spacedys.com/astdys}}. AstDyS is currently managed by the
following consortium: Department of Mathematics, University of Pisa, Italy; IASF-INAF Rome,
Italy; SpaceDyS srl, Cascina, Italy, and others. It supplies information about numbered
asteroids, with a detailed description provided by \citet{Knezevic03}. The proper orbital
elements are determined in the semimajor axis range from $1.7$~AU to $4.0$~AU, with
$464746$ asteroids in total (after removal of objects in secular resonances). For our
analysis we used proper elements derived by the synthetic method.

In Table~\ref{AstDysFamilies} we give the asteroid families detected so far in AstDys,
including the primary object, number of asteroids, and proper element ranges. See also
\citet{Milani14,Milani17}. Below we use this list as a reference for comparison with our
results. The star following the family name indicates that this family was confirmed by our
wavelet analysis (see Table~\ref{residual} below). If there are multiple stars following
the same name this means that we detected several subfamilies by wavelets.

\begin{longtable}{|c|c|cc|cc|cc|}
\caption{\label{AstDysFamilies} Currently known AstDys asteroid families.}\\
\hline
Core object & $N$ & $a_{\min}$ & $a_{\max}$  &  $e_{\min}$  & $e_{\max}$ & $\sin i_{\min}$ & $\sin i_{\max}$ \\
\hline
\endfirsthead
\multicolumn{8}{c}{\tablename\ \thetable\ -- \textit{continued}}\\
\hline
Core object & $N$ & $a_{\min}$ & $a_{\max}$  &  $e_{\min}$  & $e_{\max}$ & $\sin i_{\min}$ & $\sin i_{\max}$ \\
\hline
\endhead
\hline \multicolumn{8}{r}{\textit{Continued on next page}}\\
\endfoot
\endlastfoot
\hypertarget{h434}{434} Hungaria * & 1879 & 1.883 & 1.988 & 0.05 & 0.097 & 0.343 & 0.378\\
883 Matterania & 169 & 2.213 & 2.259 & 0.14 & 0.152 & 0.092 & 0.102\\
\hypertarget{h2076}{2076} Levin * & 1534 & 2.251 & 2.325 & 0.129 & 0.153 & 0.088 & 0.106\\
\hypertarget{h4}{4} Vesta * & 10612 & 2.256 & 2.482 & 0.08 & 0.127 & 0.1 & 0.133\\
1338 Duponta & 133 & 2.259 & 2.302 & 0.119 & 0.13 & 0.075 & 0.091 \\
298 Baptistina & 176 & 2.26 & 2.288 & 0.146 & 0.161 & 0.1 & 0.114\\
25 Phocaea & 1248 & 2.26 & 2.417 & 0.159 & 0.265 & 0.366 & 0.425 \\
135 Hertha & 15983 & 2.288 & 2.479 & 0.134 & 0.215 & 0.026 & 0.059 \\
\hypertarget{h163}{163} Erigone * & 542 & 2.331 & 2.374 & 0.2 & 0.219 & 0.08 & 0.098 \\
\hypertarget{h20}{20} Massalia * & 7820 & 2.334 & 2.474 & 0.145 & 0.175 & 0.019 & 0.034 \\
\hypertarget{h5026}{5026} Martes * & 481 & 2.368 & 2.415 & 0.2 & 0.217 & 0.082 & 0.096  \\
\hypertarget{h302}{302} Clarissa * & 236 & 2.385 & 2.421 & 0.104 & 0.111 & 0.056 & 0.06 \\
6769 Brokoff & 58 & 2.398 & 2.431 & 0.148 & 0.155 & 0.051 & 0.056 \\
752 Sulamitis & 193 & 2.42 & 2.484 & 0.084 & 0.095 & 0.085 & 0.092  \\
15 Eunomia & 9856 & 2.521 & 2.77 & 0.117 & 0.181 & 0.203 & 0.256 \\
194 Prokne & 379 & 2.522 & 2.691 & 0.154 & 0.196 & 0.292 & 0.315 \\
170 Maria & 2958 & 2.523 & 2.673 & 0.067 & 0.128 & 0.231 & 0.269  \\
480 Hansa & 1162 & 2.538 & 2.731 & 0.001 & 0.102 & 0.364 & 0.385  \\
\hypertarget{h1658}{1658} Innes * & 775 & 2.544 & 2.627 & 0.164 & 0.185 & 0.121 & 0.142 \\
3811 Karma & 59 & 2.547 & 2.579 & 0.101 & 0.11 & 0.185 & 0.19 \\
10369 Sinden & 24 & 2.551 & 2.609 & 0.104 & 0.118 & 0.469 & 0.482 \\
\hypertarget{h3815}{3815} K{\"o}nig * & 578 & 2.563 & 2.584 & 0.138 & 0.143 & 0.144 & 0.164  \\
\hypertarget{h606}{606} Brang{\"a}ne * & 325 & 2.571 & 2.597 & 0.178 & 0.183 & 0.165 & 0.168  \\
\hypertarget{h145}{145} Adeona * & 2070 & 2.573 & 2.714 & 0.153 & 0.181 & 0.193 & 0.213 \\
4203 Brucato & 41 & 2.586 & 2.69 & 0.119 & 0.138 & 0.47 & 0.489  \\
945 Barcelona & 346 & 2.591 & 2.668 & 0.189 & 0.289 & 0.506 & 0.521  \\
116763 & 24 & 2.612 & 2.652 & 0.236 & 0.246 & 0.463 & 0.468\\
\hypertarget{h3}{3} Juno * & 1693 & 2.622 & 2.7 & 0.227 & 0.245 & 0.225 & 0.239  \\
569 Misa & 647 & 2.623 & 2.694 & 0.169 & 0.184 & 0.034 & 0.045 \\
7744 & 98 & 2.633 & 2.67 & 0.069 & 0.075 & 0.041 & 0.049 \\
\hypertarget{h1547}{1547} Nele * & 344 & 2.638 & 2.65 & 0.266 & 0.27 & 0.21 & 0.213 \\
29841 & 65 & 2.639 & 2.668 & 0.052 & 0.059 & 0.033 & 0.04  \\
17392 & 96 & 2.645 & 2.681 & 0.059 & 0.07 & 0.036 & 0.042 \\
23255 & 12 & 2.655 & 2.7 & 0.095 & 0.113 & 0.46 & 0.469 \\
2782 Leonidas & 111 & 2.657 & 2.701 & 0.185 & 0.197 & 0.06 & 0.072 \\
10955 Harig & 918 & 2.671 & 2.762 & 0.005 & 0.026 & 0.1 & 0.113 \\
12739 & 298 & 2.682 & 2.746 & 0.047 & 0.06 & 0.031 & 0.041 \\
11882 & 87 & 2.683 & 2.711 & 0.059 & 0.066 & 0.031 & 0.04 \\
110 Lydia & 898 & 2.696 & 2.779 & 0.026 & 0.061 & 0.083 & 0.106  \\
\hypertarget{h808}{808} Merxia ** & 1263 & 2.705 & 2.81 & 0.125 & 0.143 & 0.08 & 0.093  \\
410 Chloris & 120 & 2.705 & 2.761 & 0.238 & 0.266 & 0.146 & 0.16  \\
\hypertarget{h3827}{3827} Zden\v{e}khorsk{\'y} * & 1050 & 2.705 & 2.768 & 0.082 & 0.096 & 0.08 & 0.094 \\
21344 & 75 & 2.708 & 2.741 & 0.15 & 0.16 & 0.046 & 0.05 \\
53546 & 81 & 2.709 & 2.735 & 0.169 & 0.174 & 0.247 & 0.251  \\
14916 & 17 & 2.71 & 2.761 & 0.27 & 0.282 & 0.537 & 0.542  \\
148 Gallia & 137 & 2.71 & 2.812 & 0.114 & 0.15 & 0.42 & 0.43 \\
\hypertarget{h847}{847} Agnia ** & 3336 & 2.713 & 2.819 & 0.063 & 0.083 & 0.055 & 0.076  \\
40134 & 16 & 2.715 & 2.744 & 0.223 & 0.235 & 0.429 & 0.44 \\
\hypertarget{h93}{93} Minerva * & 2428 & 2.718 & 2.816 & 0.115 & 0.155 & 0.146 & 0.169\\
729 Watsonia & 83 & 2.72 & 2.816 & 0.11 & 0.144 & 0.294 & 0.305  \\
\hypertarget{h396}{396} Aeolia * & 529 & 2.728 & 2.752 & 0.163 & 0.171 & 0.057 & 0.062 \\
\hypertarget{h668}{668} Dora * & 1742 & 2.744 & 2.812 & 0.188 & 0.204 & 0.128 & 0.143 \\
2 Pallas & 45 & 2.752 & 2.791 & 0.254 & 0.283 & 0.531 & 0.55 \\
\hypertarget{h1128}{1128} Astrid * & 548 & 2.754 & 2.817 & 0.045 & 0.053 & 0.008 & 0.019 \\
\hypertarget{h1726}{1726} Hoffmeister * & 2095 & 2.754 & 2.82 & 0.041 & 0.053 & 0.066 & 0.088 \\
13314 & 241 & 2.756 & 2.804 & 0.17 & 0.183 & 0.069 & 0.079 \\
\hypertarget{h18466}{18466} * & 257 & 2.763 & 2.804 & 0.171 & 0.182 & 0.229 & 0.236 \\
32418 & 81 & 2.763 & 2.795 & 0.255 & 0.261 & 0.152 & 0.156 \\
1222 Tina & 107 & 2.764 & 2.811 & 0.065 & 0.113 & 0.349 & 0.36 \\
\hypertarget{h158}{158} Koronis **** & 7390 & 2.816 & 2.985 & 0.016 & 0.101 & 0.029 & 0.047  \\
\hypertarget{h293}{293} Brasilia * & 845 & 2.832 & 2.874 & 0.118 & 0.133 & 0.256 & 0.264 \\
18405 & 159 & 2.832 & 2.859 & 0.103 & 0.11 & 0.158 & 0.162 \\
\hypertarget{h16286}{16286} *& 94 & 2.846 & 2.879 & 0.038 & 0.047 & 0.101 & 0.111 \\
1189 Terentia & 80 & 2.904 & 2.936 & 0.07 & 0.075 & 0.192 & 0.194  \\
\hypertarget{h845}{845} Na{\"e}ma * & 375 & 2.914 & 2.962 & 0.029 & 0.041 & 0.205 & 0.209 \\
\hypertarget{h179}{179} Klytaemnestra *& 513 & 2.946 & 3.015 & 0.051 & 0.081 & 0.147 & 0.16  \\
\hypertarget{h221}{221} Eos * & 16038 & 2.948 & 3.211 & 0.022 & 0.133 & 0.148 & 0.212  \\
\hypertarget{h283}{283} Emma * & 577 & 3.028 & 3.086 & 0.107 & 0.124 & 0.154  & 0.166 \\
7468 Anfimov & 49 & 3.031 & 3.075 & 0.087 & 0.091 & 0.059  & 0.061 \\
3438 Inarradas & 43 & 3.036 & 3.076 & 0.174 & 0.186 & 0.249 & 0.255 \\
\hypertarget{h96}{96} Aegle * & 120 & 3.036 & 3.083 & 0.176 & 0.189 & 0.279 & 0.289  \\
\hypertarget{h24}{24} Themis ** & 5612 & 3.062 & 3.24 & 0.114 & 0.192 & 0.009 & 0.049 \\
10 Hygiea & 3145 & 3.067 & 3.242 & 0.1 & 0.166 & 0.073 & 0.106  \\
21885 & 61 & 3.079 & 3.112 & 0.025 & 0.035 & 0.184 & 0.189  \\
31 Euphrosyne & 1385 & 3.082 & 3.225 & 0.149 & 0.231 & 0.431 & 0.459 \\
\hypertarget{h1040}{1040} Klumpkea * & 1815 & 3.083 & 3.175 & 0.176 & 0.217 & 0.279 & 0.304 \\
780 Armenia & 67 & 3.085 & 3.133 & 0.06 & 0.075 & 0.31 & 0.314 \\
1298 Nocturna & 186 & 3.088 & 3.22 & 0.105 & 0.124 & 0.103 & 0.125\\
159 Aemilia & 62 & 3.091 & 3.131 & 0.111 & 0.117 & 0.084 & 0.09 \\
31811 & 144 & 3.095 & 3.14 & 0.059 & 0.075 & 0.178 & 0.188  \\
375 Ursula & 731 & 3.095 & 3.241 & 0.057 & 0.13 & 0.264 & 0.303 \\
5651 Traversa & 56 & 3.097 & 3.166 & 0.111 & 0.129 & 0.231 & 0.241  \\
43176 & 75 & 3.107 & 3.156 & 0.065 & 0.075 & 0.174 & 0.184 \\
58892 & 20 & 3.113 & 3.154 & 0.152 & 0.163 & 0.3 & 0.308 \\
8737 Takehiro & 57 & 3.116 & 3.143 & 0.112 & 0.121 & 0.207 & 0.211\\
\hypertarget{h3330}{3330} Gantrisch * & 1241 & 3.118 & 3.178 & 0.184 & 0.213 & 0.171 & 0.184 \\
1118 Hanskya & 116 & 3.133 & 3.249 & 0.034 & 0.059 & 0.252 & 0.267  \\
22805 & 20 & 3.135 & 3.167 & 0.165 & 0.175 & 0.301 & 0.308  \\
\hypertarget{h490}{490} Veritas * & 2139 & 3.143 & 3.197 & 0.048 & 0.08 & 0.151 & 0.173\\
7605 & 19 & 3.143 & 3.154 & 0.063 & 0.075 & 0.447 & 0.453 \\
\hypertarget{h778}{778} Theobalda *& 574 & 3.155 & 3.199 & 0.239 & 0.261 & 0.243 & 0.253 \\
3460 Ashkova & 59 & 3.159 & 3.219 & 0.186 & 0.211 & 0.016 & 0.028  \\
5931 Zhvanetskij & 23 & 3.174 & 3.215 & 0.16 & 0.172 & 0.302 & 0.313  \\
618 Elfriede & 97 & 3.177 & 3.2 & 0.056 & 0.059 & 0.27 & 0.278  \\
6355 Univermoscow & 13 & 3.188 & 3.217 & 0.088 & 0.097 & 0.374 & 0.378  \\
3025 Higson & 17 & 3.188 & 3.221 & 0.059 & 0.066 & 0.366 & 0.378 \\
1303 Luthera & 232 & 3.192 & 3.237 & 0.106 & 0.144 & 0.31 & 0.337 \\
895 Helio & 50 & 3.194 & 3.225 & 0.168 & 0.183 & 0.437 & 0.446\\
69559 & 17 & 3.201 & 3.219 & 0.196 & 0.201 & 0.299 & 0.305\\
10654 Bontekoe & 13 & 3.207 & 3.244 & 0.051 & 0.056 & 0.368 & 0.374\\
1101 Clematis & 17 & 3.229 & 3.251 & 0.03 & 0.037 & 0.363 & 0.375 \\
11097 & 33 & 3.274 & 3.275 & 0.23 & 0.28 & 0.014 & 0.038 \\
45637 & 20 & 3.341 & 3.369 & 0.103 & 0.123 & 0.142 & 0.151 \\
260 Huberta & 26 & 3.41 & 3.464 & 0.079 & 0.089 & 0.099 & 0.108 \\
87 Sylvia & 191 & 3.458 & 3.567 & 0.046 & 0.074 & 0.162 & 0.179  \\
909 Ulla & 37 & 3.524 & 3.568 & 0.043 & 0.058 & 0.306 & 0.309  \\
3561 Devine & 19 & 3.962 & 3.962 & 0.127 & 0.133 & 0.149 & 0.156 \\
1911 Schubart & 531 & 3.964 & 3.967 & 0.158 & 0.224 & 0.039 & 0.056 \\
153 Hilda & 18 & 3.965 & 3.966 & 0.171 & 0.181 & 0.152 & 0.156\\
6124 Mecklenburg & 78 & 3.966 & 3.967 & 0.186 & 0.212 & 0.146 & 0.159 \\
\hline
\end{longtable}

\section{Statistical wavelet analysis}
\label{sec_statwa}
We emphasize that in this work the notion ``asteroid family'' is understood in the
statistical sense, that is as a group of objects that are statistically unlikely to
originate from the randomness of the asteroid sample (from its shot noise). Therefore, such
a group should be generated by some physical mechanism, but this does not necessarily
suggests that these objects have common origin.

A self-consistent method perform wavelet analysis was presented in \citep{Baluev18a}. That
first release handled several theoretic issues that had not a good solution before, and
also involved wavelets of optimized shape, but that analysis tool targeted only 1D
distributions. However, the first test application of this initial 1D method to
exoplanetary population revealed several rather interesting results \citep{Baluev18b}. In
particular, hints of a previously unknown subtle family of giant exoplanets was detected,
possibly related to the iceline accumulation effect in a protoplanetary disk. Finally, a 2D
generalization of this algorithm was presented in \citep{BalRodShai19}. This analysis is
based on isotropic radially symmetric wavelets, so it requires that two input parameters in
the 2D sample are at least physically and numerically comparable (like e.g. two coordinates
in the Euclidean space).

When processing exoplanetary samples that contained $\sim 10^3$ objects at most, it was not
possible to derive entirely reliable and convincing results. For example, the formal
significance of the exoplanetary family mentioned above was in the range $2-3$ sigma, i.e.
its interpretation is still probabilistic. However, the Main Belt asteroids form a much
larger sample containing between $10^5$ and $10^6$ objects. This should allow much more
statistically reliable conclusions and more convincing detections.

For the full mathematical details of the wavelet analysis algorithm the reader is referred
to \citep{Baluev18a} and \citep{BalRodShai19}. Here we omit these details, only giving a
few general ideas of the method.

First of all, the continuous wavelet transform (or CWT) is defined as
\begin{equation}
Y(s,b) = \int\limits_{\mathbb R^n} f(x) \psi\left(\frac{x-b}{s}\right)\, dx.
\label{cwt}
\end{equation}
This assumes a general $n$-dimensional task, where $n=1$ or $n=2$ refers to the dimension
of $x$ and of the shift parameter $b$. The scale parameter $s$ is always scalar here. The
wavelet $\psi$ is assumed radially-symmetric for $n=2$, so it is actually a function of the
length of its vector argument. Notice that the scale parameter is usually denoted as $a$,
but we changed it to $s$ in order to avoid mixing it with the semimajor axis below.

The CWT can be inverted using the following general inversion formula:
\begin{equation}
f(x) = \frac{1}{C_{\psi\gamma}} \int\limits_0^{\infty} \frac{ds}{s^{2n+1}} \int\limits_{\mathbb R^n} Y(s,b) \gamma\left(\frac{x-b}{s}\right)\, db,
\label{icwt}
\end{equation}
where $\gamma$ is a largerly arbitrary inversion kernel. Although the most popular version
of~(\ref{icwt}) involves $\gamma=\psi$, it is possible to consider other $\gamma$ as well,
resulting in certain useful specific properties of the inversion formula. The constant
$C_{\psi\gamma}$ in~(\ref{icwt}) depends solely on the choice of $\psi$ and $\gamma$.

\begin{figure}
\begin{center}
\includegraphics[width=\textwidth]{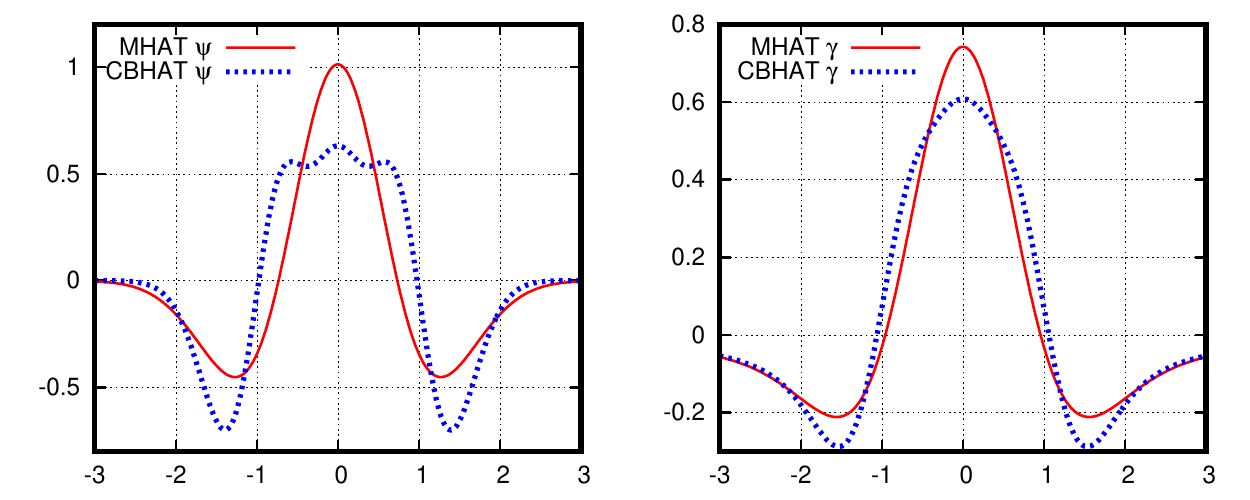}\\
\includegraphics[width=\textwidth]{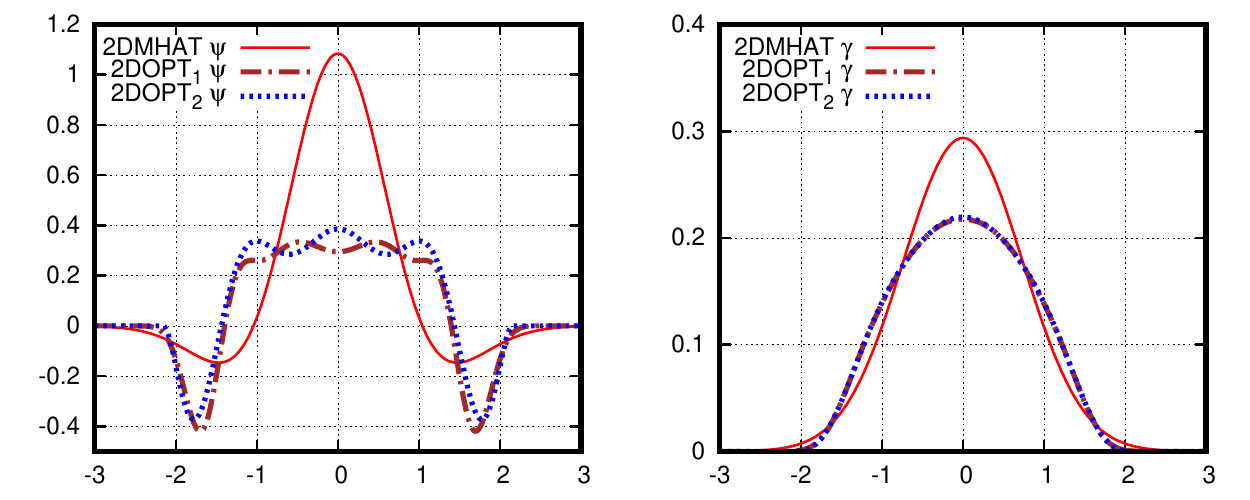}
\caption{
\label{wavs}
     The optimized wavelet function $\psi$ for the CBHAT and 2DOPT wavelets and their
     optimal reconstruction kernels $\gamma$, compared to the classic 1D and 2D MHAT
     wavelets with their optimal $\gamma$.}
\end{center}
\end{figure}

We shape of the wavelet function $\psi$ is rather arbitrary and largerly depends on the
goals of the analysis. In our work we will use optimized wavelets that allow to
simultaneously minimize the noise level and improve the noise gaussianity. Such 1D and 2D
wavelets, named as CBHAT (`cowboy hat') and 2DOPT, were derived in
\citep{Baluev18a,BalRodShai19}, and are plotted in Fig.~\ref{wavs}. Notice that they are
different from the famous classic MHAT wavelet that does not suit in our task because it
generates bad noise properties. Notice that we have two very close versions of 2DOPT, among
which we select 2DOPT$_2$ (and thus omit this index hereafter). All these wavelets are such
that $Y(s,b)$ represents a smoothed second derivative or a smoothed Laplacian of $f(x)$,
with smoothing scale controlled by $s$. In Fig.~\ref{wavs} we also plot optimized inversion
kernels $\gamma$ that allow to minimize the noise in the reconstructed $f(x)$.

We aim to apply the CWT to the 1D or 2D probability density function p.d.f. $f(x)$.
However, this function is not observed directly, so the formulae~(\ref{cwt})
and~(\ref{icwt}) cannot be used at this stage. What we have in practice in place of $f(x)$
is the sample $\{x_i\}_{i=1}^N$, and hence we may only construct a statistical estimate for
the CWT:
\begin{equation}
\tilde Y(s,b) = \frac{1}{N} \sum_{i=1}^N \psi\left(\frac{x_i-b}{s}\right).
\label{swt}
\end{equation}
Notice that the CWT itself, as defined in~(\ref{cwt}), is a mathematical expectation of
$y=\psi[(x-b)/s]$, where $s$ and $b$ are parameters, so in~(\ref{swt}) involves plainly
corresponding sample mean of the same $y$. We therefore refer to~(\ref{swt}) as to sample
wavelet transform (SWT).

This is the point where the noise appears. The SWT is a noisy quantity since it is defined
on the basis of a finite sample. It is easy to define the sample variance $\tilde D$ in the
way similar to~(\ref{swt}). Finally, we can construct the normalized test statistic
\begin{equation}
z(s,b) = \frac{\tilde Y(s,b) - Y(s,b)}{\sqrt{\tilde D(s,b)}},
\end{equation}
which has asymptotically (for large $N$) the standard Gaussian distribution (mean zero,
variance unit). Notice that we can substitute here any comparison model $Y_0$ in place of
$Y$. Basically, our formal goal is to test whether some null hypothesis $Y=Y_0$ is
statistically consistent or not.

The test statistic $z(s,b)$ is the central testing quantity that allows to derive whether
the wavelet coefficient (the value of $\tilde Y$) is statistically sound at the given
$(s,b)$. The typical noise would imply $z$ of the order of unit, while a large $z$
indicates a statistically significant inconsistency between the adopted comparison model
$Y_0$ and the actual sample distribution. The Gaussian asymptotic distribution of $z$ can
be used to construct a formal statistical test.

However, the reader is cautioned that it is inadequate to apply such approach literally if
multiple $(s,b)$ points are tested (which is typically the case). We usually investigate a
wide domain $\mathcal D$ in the $(s,b)$-plane, so the actual compound test basically
involves multiple elementary tests per independent $z$-values. In such a case it is
mandatory to apply some statistical correction for multiple testing. We put a special
emphasis on this issue because it was often ignored so far in many other works, thus
resulting in a drastically increased level of false positives among the detected wavelet
coefficients.

In our framework, the multiple testing issue can be handled neatly if we consider the
extreme value statistic instead of the single-value ones. Namely, what we test in actuality
is the maximum deviation
\begin{equation}
z_{\max} = \max_{(s,b)\in \mathcal D} |z(s,b)|
\end{equation}
instead of the particular $z(s,b)$ values. The distribution function of $z_{\max}$ is
non-Gaussian, but it can be characterized analytically as an extreme value distribution of
a Gaussian random field $z(s,b)$. This work was done in \citep{Baluev18a,Baluev18b},
resulting in the following tail approximation
\begin{equation}
\FAP(\zeta) \equiv \Pr\{z_{\max}>\zeta\} \lesssim 2 W_{00} \zeta^n e^{-\frac{\zeta^2}{2}}.
\label{fap}
\end{equation}
This formula connects the false alarm probability (FAP) with the maximum observed
$z$-level. If the resulting $\FAP(z_{\max})$, computed for the actually observed
$z_{\max}$, is smaller than a conventional threshold level (say, $1$ per cent or any) then
the deviation is treated significant and the comparison model $Y_0$ disagrees with the
sample. The coefficient $W_{00}$ depends on the wavelet $\psi$ and on the domain $\mathcal
D$, and it can be computed numerically together with the SWT. Importantly,
formula~(\ref{fap}) has the shape of an approximate upper bound, so its possible
inaccuracies should not lead to understated $\FAP$ (overstated significance). If the right
hand side of~(\ref{fap}) is below some $\FAP_{\rm thr}$ than the actual $\FAP$ is also
below than this threshold.

Concerning the domain $\mathcal D$, it can be chosen rather arbitrary. In fact, it
accumulates our prior assumptions, where we expect to find a signfificant wavelet
coefficient, and where not. However, this domain cannot be arbitrarily large. In any case,
it should be restricted to the domain where $z(s,b)$ is satisfactorily Gaussian, because it
was our substantial assumption used to compute the FAP approximation~(\ref{fap}). We
typically expand $\mathcal D$ to this widest range, while the normality is verified using
certain formalized criterion \citep{Baluev18a,BalRodShai19}.

Our statistical test based on~(\ref{fap}) only allows to decide whether some given
comparison model $f_0/Y_0$ agrees with the sample or not. However, this might appear not
enough for our goals, because we would also like to learn, how the p.d.f. $f(x)$ should
look to satisfy this restriction. In other word, we should construct some most economic
p.d.f. model not violating the significance test. This is achieved through an iterative
scheme with a single iteration layed out below:
\begin{equation}
f_n(x) \stackrel{\rm{CWT}}{\longmapsto} Y_n(s,b) \stackrel{\rm{noise\ thresholding}}{\longmapsto} Y_{n,\rm thr}(s,b) \stackrel{\rm{CWT}^{-1}}{\longmapsto} f_{n+1}(x).
\label{ftf}
\end{equation}
Here, the noise thresholding stage is performed based on the significance thresholds $z_{\rm
thr}$ derived from~(\ref{fap}). This is basically a matching pursuit algorithm that allows
to construct the p.d.f. model in the most economic manner, i.e. by using the smallest
possible number of nonzero wavelet coefficients, simultaneously satisfying the test
condition for $\FAP(z_{\max})$.

Further details can be found in \citep{Baluev18a,BalRodShai19}. The code is available for
download at \url{https://sourceforge.net/projects/waveletstat/}. The computations in this
work were done using an Intel Core i9 9900K workstation with 64~Gb of memory.

\section{Analysis of 1D distributions}
\label{sec_1D}
For each of the 1D distribution considered below, we plot two graphs: the 2D significance
map $g(z(s,b))$ corresponding to the very first step of the iterative process~(\ref{ftf}),
and the 1D reconstructed p.d.f. model obtained after all the iterations~(\ref{ftf}).

The 2D significance map $g(z(s,b))$ is formally defined in \citep{Baluev18b}. In brief,
each value in such a map represents a normal quantile for $z(s,b)$, i.e. the significance
of the given $z$-value, as would be expressed in terms of Gaussian standard deviations. For
example, $g=2$ means the two-sigma significance (FAP about $5\%$), $g=3$ is three-sigma
(FAP about $0.27\%$), and so on. The higher is $g$, the more statistically sound is the
wavelet coefficient corresponding to the given point $(s,b)$. The points in the
significance map with $g<1$ are entirely insignificant, and are always rendered as white.
Formally, $g$ would always be non-negative, but we conventionally define it signed,
assuming that $g<0$ means $z<0$. Further guidelines on how to interpret the 2D significance
maps plotted below can be found in \citep{Baluev18b}, along with several tutorial cases and
cautions.

In the 2D maps we show only the domains where $z(s,b)$ has near-Gaussian distribution. The
non-Gaussian domains, where the results cannot be trusted, are hashed out by gray. Also,
the 2D graphs contain a black line in the bottom (small-scale range) which represents the
Gaussian domain boundary, as computed using an approximate formula.

In the 1D graphs, the reconstructed p.d.f. models $\tilde f(x)$ are plotted for three
significance thresholds, corresponding to 1-sigma, 2-sigma, and 3-sigma levels. However, in
this work all them appeared practically identical, again because of the sharp transition
between significant and insignificant domains in the 2D significance maps.

The matching pursuit iterations always started from the best fitting Gaussian distribution
$\tilde f_0(x) = f_G(x)$ (i.e., the significance map refers to the difference $f-f_G$).
This is a bit different from \citep{Baluev18b}, where they started from $f_0 \equiv 0$.

\subsection{Distributions of physical parameters}
We first considered several physical asteroid parameters: diameter, absolute magnitude, and
color index ($B-V$). The first two distributions appeared simply unimodal without any
details, so they are omitted.

\begin{figure}
\begin{center}
\includegraphics[width=\textwidth]{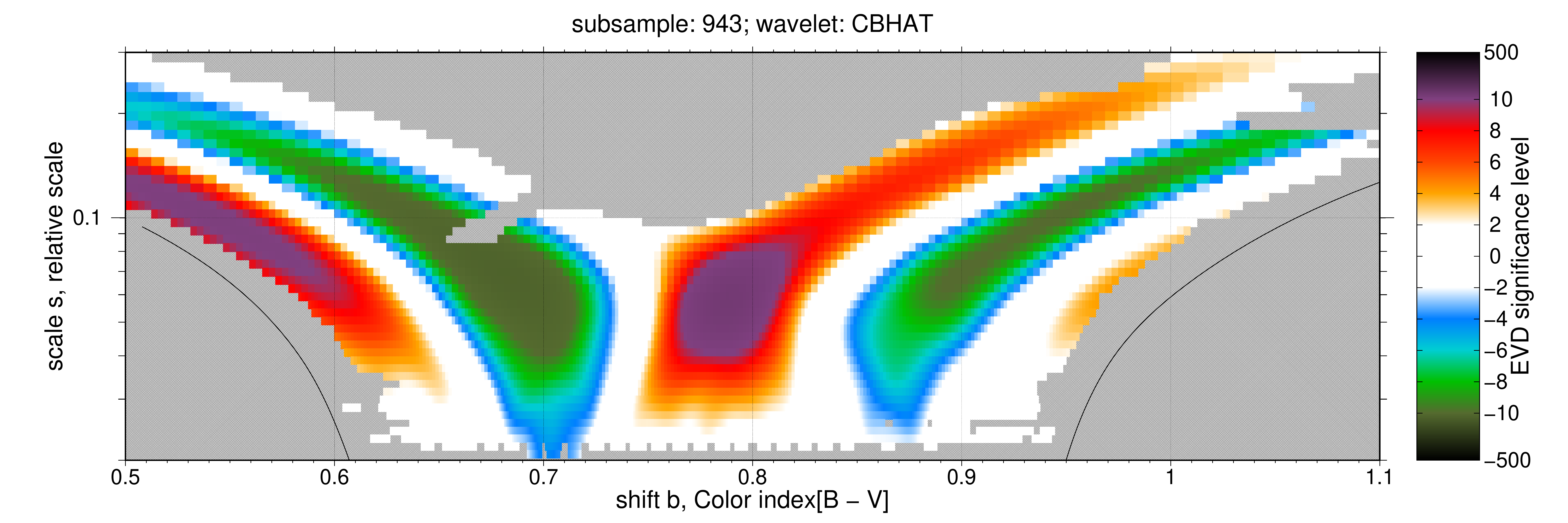}\\
\includegraphics[width=\textwidth]{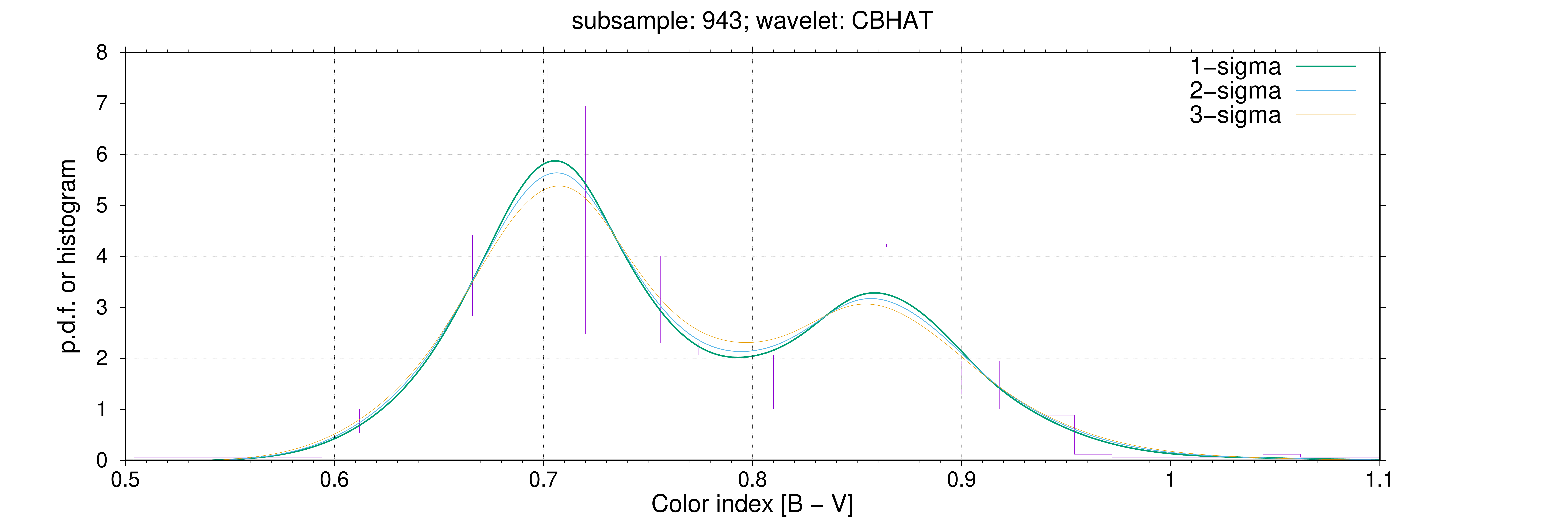}
\caption{
\label{BV}
     Wavelet analysis for the distribution of asteroids color index $B-V$, CBHAT.}
\end{center}
\end{figure}

The color index $B-V$ appeared more interesting, shown in Fig.~\ref{BV}. It reveals a
bimodality with a clear gap between two modes, near $0.71$ and $0.86$. The larger peak is
likely related to carbonaceous asteroids, while the smaller peak contains rocky asteroids.

These distributions of physical parameters appeared quite simple. We were able to resolve
only the large-scale patterns that could be easily seen in histograms. The wavelet analysis
only confirmed that there are no detectable small-scale details.

\subsection{Distributions of proper orbital elements}
Finally, we proceed to the proper orbital elements. Now we consider the same three orbital
parameters $e$, $i$ and $a$, as in the osculating case.

\begin{figure}
\begin{center}
\includegraphics[width=\textwidth]{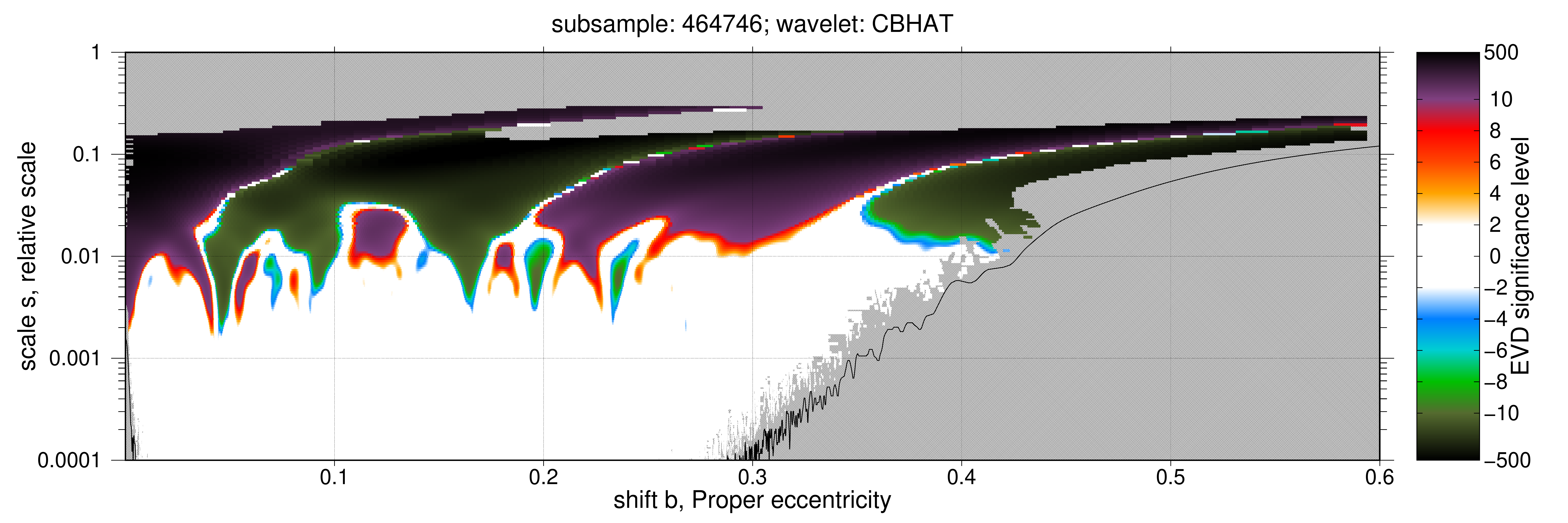}\\
\includegraphics[width=\textwidth]{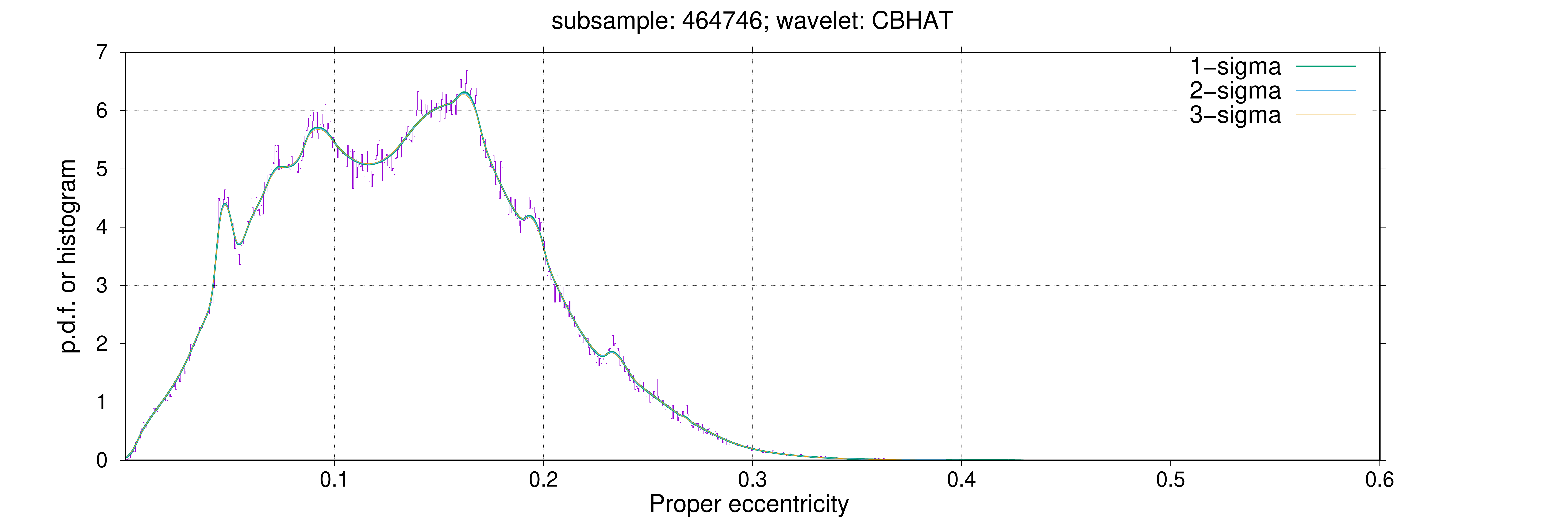}
\caption{
\label{proper-ecc}
     Wavelet analysis for the distribution of asteroids proper orbital eccentricity $e_p$, CBHAT.}
\end{center}
\end{figure}

As we can see from Fig.~\ref{proper-ecc}, the distribution of proper eccentricities
demonstrates multiple local inhomogeneities. Those inhomogeneities are likely related to
various asteroid families. For example, the density concentration for $e_p$ in the range
$0.045-0.05$ is possibly related to the Hoffmeister and Astrid families, the range
$0.19-0.2$ is related to Dora family (see Table~\ref{AstDysFamilies}). However it is not
easy to set a one-to-one correspondence between families from Table~\ref{AstDysFamilies}
and peaks of the 1D distribution of $e_p$. This is probably because multiple families
overlap with each other in such 1D view.

\begin{figure}
\begin{center}
\includegraphics[width=\textwidth]{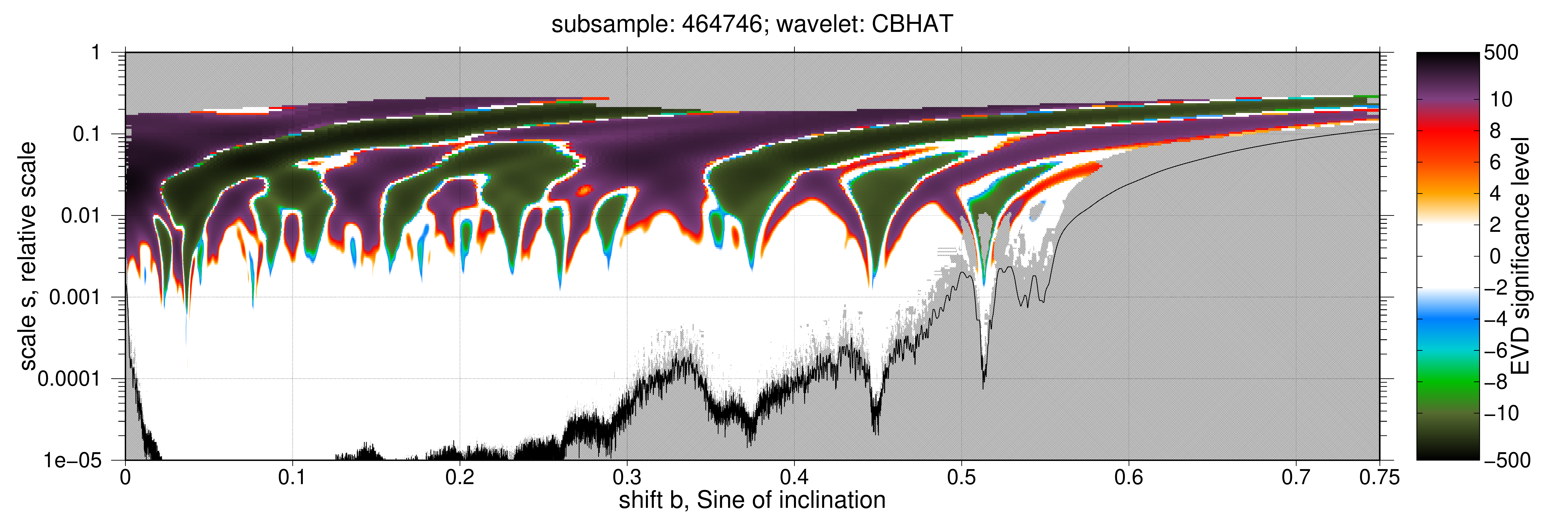}\\
\includegraphics[width=\textwidth]{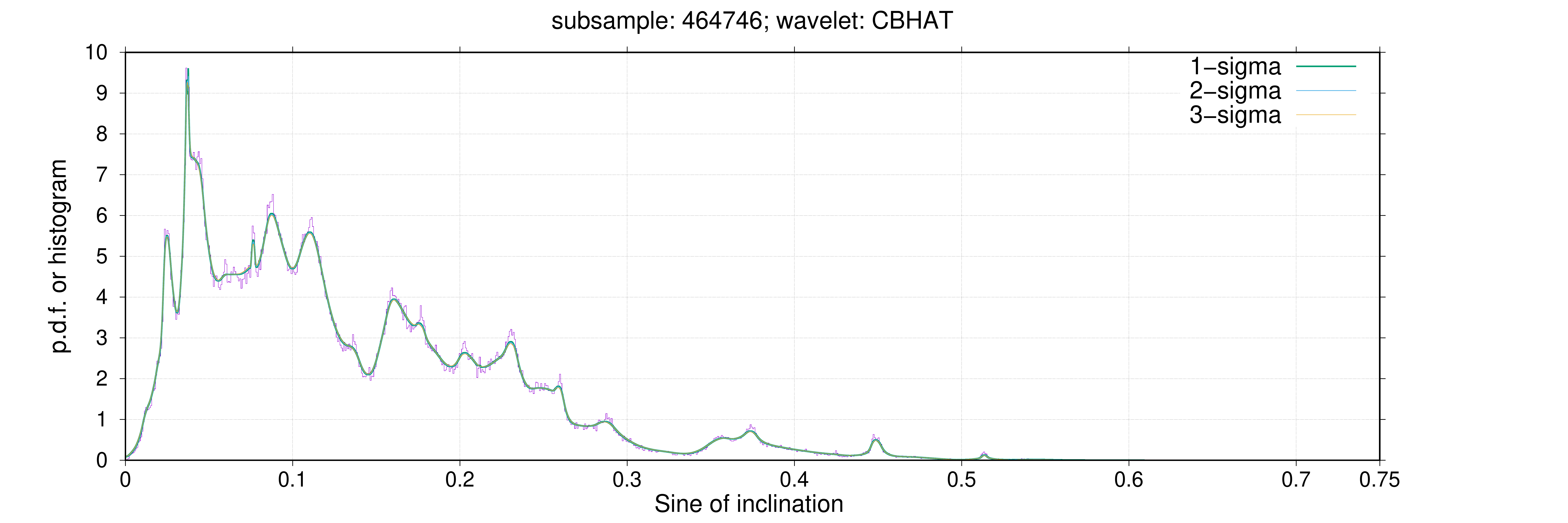}
\caption{
\label{proper-inc}
     Wavelet analysis for the distribution of asteroids proper inclination $i_p$, CBHAT.}
\end{center}
\end{figure}

The proper orbital inclination (Fig.~\ref{proper-inc}) reveals qualitatively similar
behaviour. At least $15$ local concentrations can be detected, which can be related to the
asteroid families, or some dynamical effects. However, it is again difficult to
unambiguously separate these families from each other based on just the 1D analysis.

\begin{figure}
\begin{center}
\includegraphics[width=\textwidth]{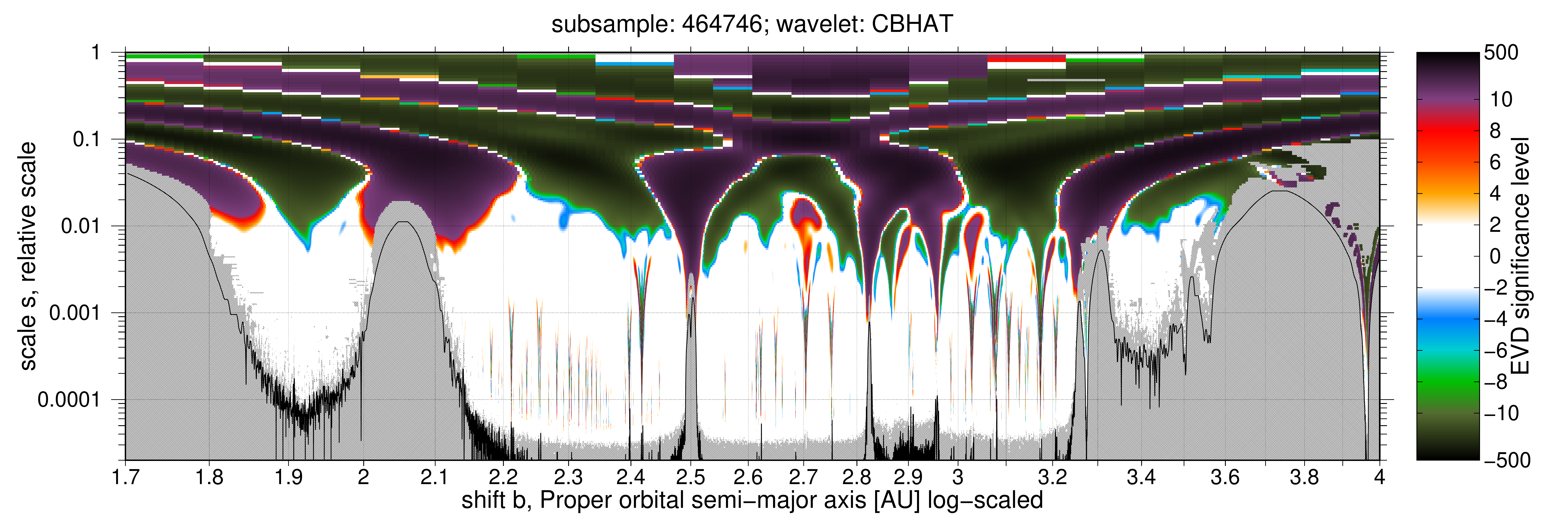}\\
\includegraphics[width=\textwidth]{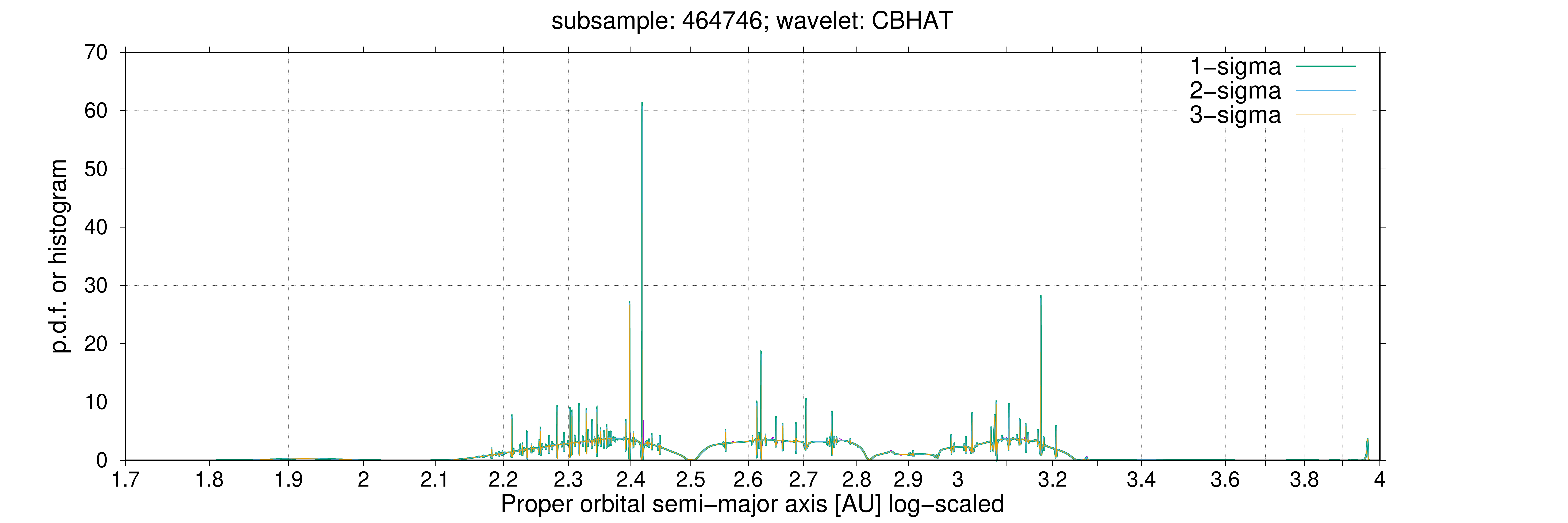}\\
\includegraphics[width=\textwidth]{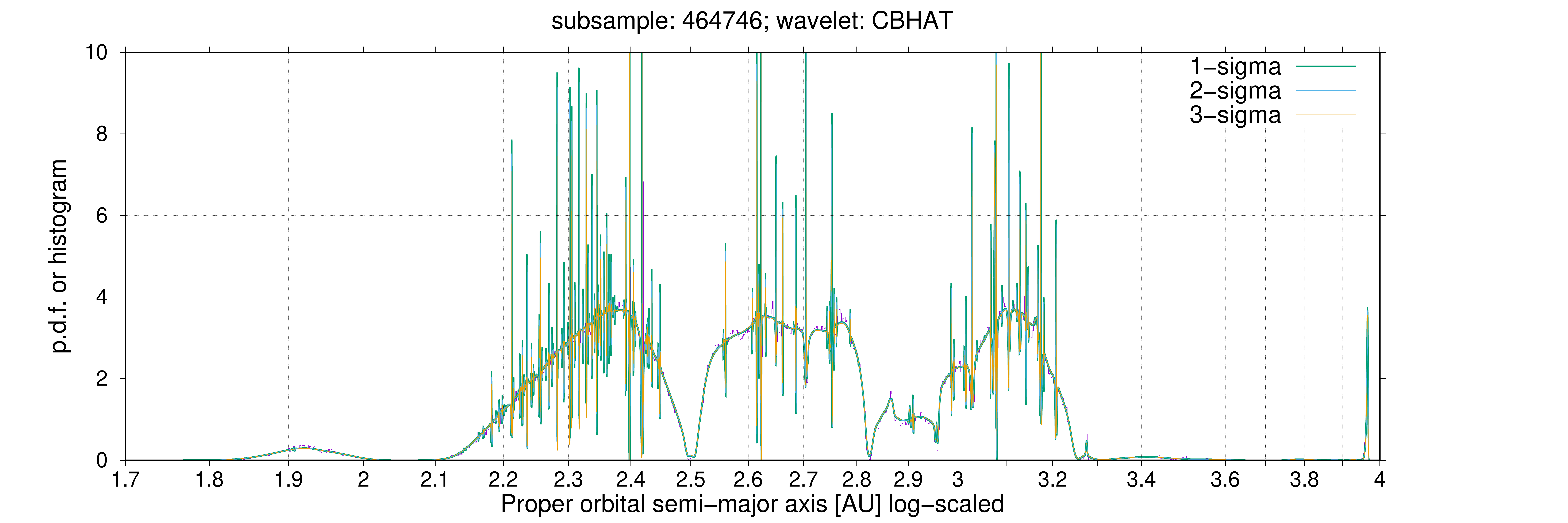}
\caption{
\label{proper-smaxes}
     Wavelet analysis for the distribution of asteroids proper semimajor axis $a_p$, CBHAT.
     The second p.d.f. plot has a cutted ordinate to show the most informative portion of the graph.}
\end{center}
\end{figure}

The distribution of the proper semimajor axis (Fig.~\ref{proper-smaxes}) appears the most
informative and the most interesting among all other 1D distributions. The thin resonant
bands (gaps as well as concentrations) are detected very easily. However, such extremely
narrow groups are mainly associated to just the mean-motion resonaces affecting the motion
of the asteroids. They are not related to the ``asteroid families'' in the genetic sense of
this notion. We revealed $110$ such resonant asteroid groups, they are given in
Table~\ref{ResonantFamilies}. In the first column we show the number of the brightest
asteroid of a group (or the smallest absolute magnitude).

Notice that although we attribute them to resonances here, and some of them indeed have
obvious commensurability with e.g. Jupiter, we did not formally verify that the resonant
dynamics indeed takes place.

\begin{figure}
\begin{center}
\includegraphics[width=\textwidth]{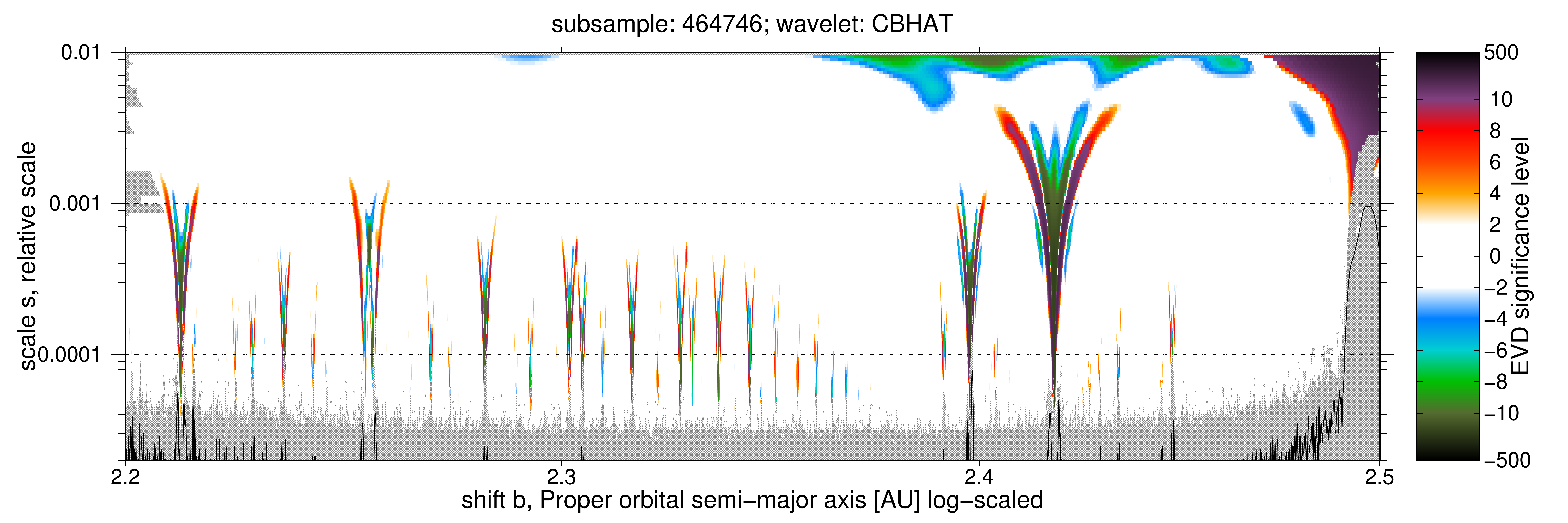}\\
\includegraphics[width=\textwidth]{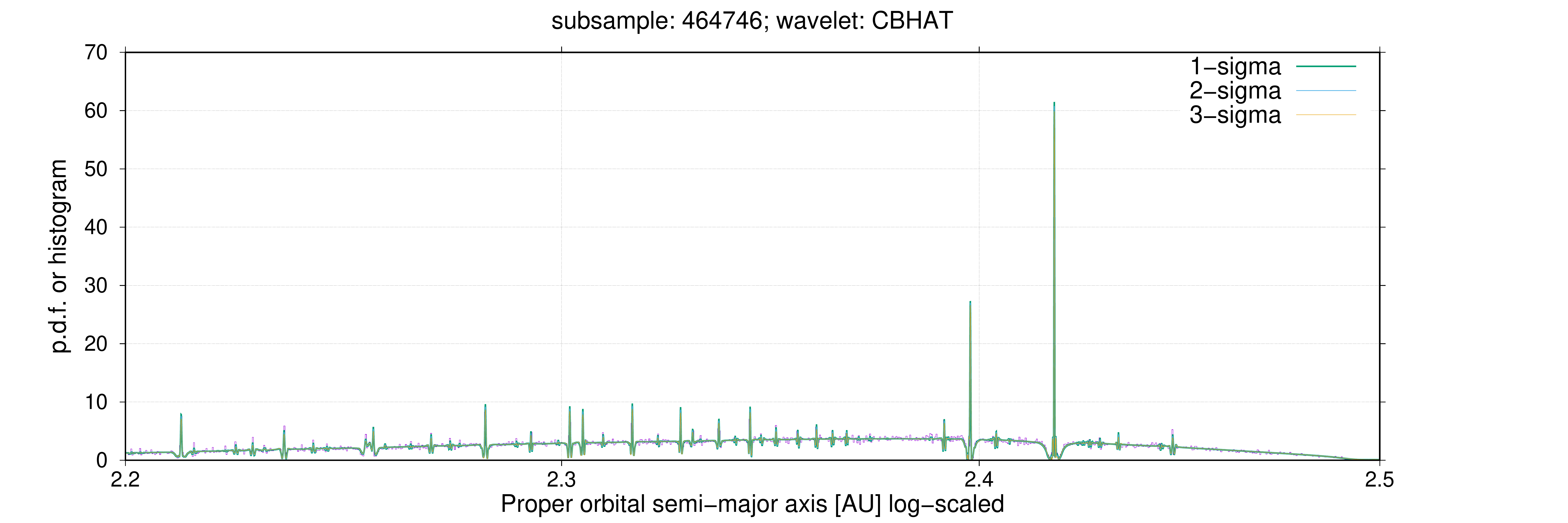}\\
\includegraphics[width=\textwidth]{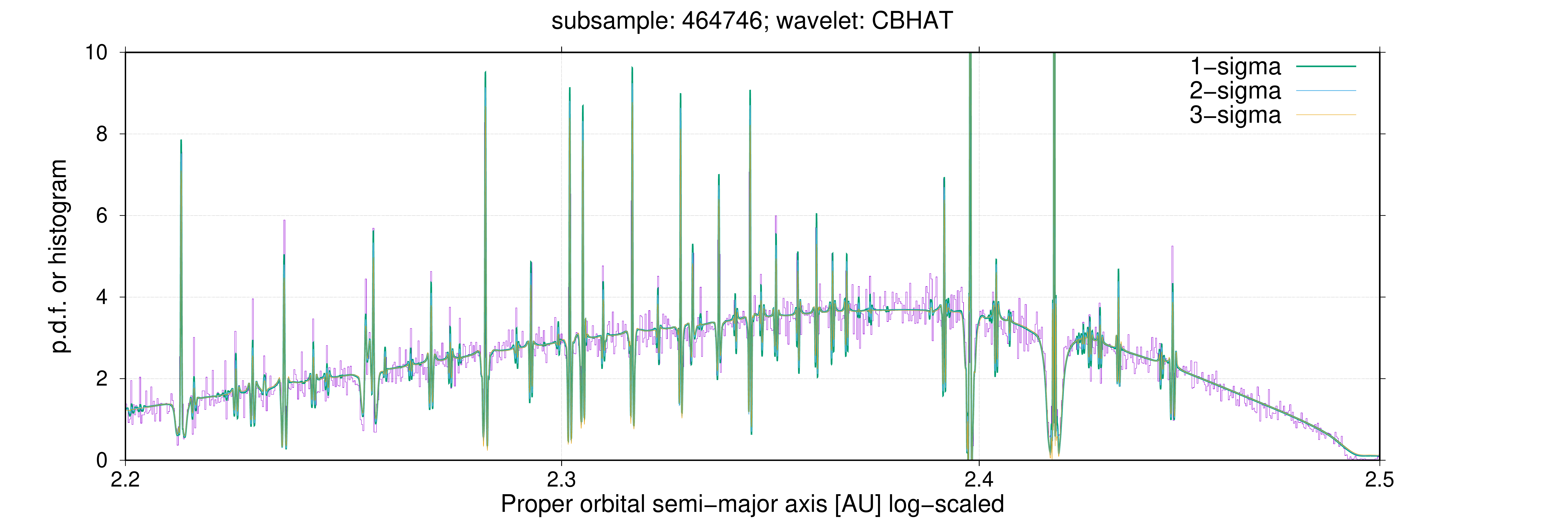}
\caption{
\label{proper-smaxes-cut}
     Wavelet analysis for the distribution of asteroids proper semimajor axis $A_p$, range $2.2-2.5$~AU, CBHAT.
     The second p.d.f. plot has a cutted ordinate to show the most informative portion of the graph.}
\end{center}
\end{figure}

For a more clear presentation we also plot in Fig.~\ref{proper-smaxes-cut} an expanded
small portion of Fig.~\ref{proper-smaxes} in the cutted range $2.2-2.5$~AU.

In addition we may notice that our 1D analysis is capable to easily resolve the internal
structure of the resonant families, and this fine structure appears rather intricate. Each
such family has an extremely thin core surrounded by two wider gaps from the both sides.
Moreover, the shape of the core appears very peaky, relatively to e.g. the Gaussian bell
shape. This might be interrelated with some properties of resonant motion, or with
artifacts of the averaging procedure used to derive proper semimajor axis.

\begin{longtable}{|c|c|cc|c|c|c|cc|}
\caption{\label{ResonantFamilies} Resonant asteroid families detected by the 1D analysis.}\\
\cline{1-4}
\cline{6-9}
Core & $N$ & $a_{\min}$ & $a_{\max}$ &~~~~~& Core & $N$ & $a_{\min}$ & $a_{\max}$ \\
\cline{1-4}
\cline{6-9}
\endfirsthead
\multicolumn{9}{c}{\tablename\ \thetable\ -- \textit{continued}}\\
\cline{1-4}
\cline{6-9}
Core & $N$ & $a_{\min}$ & $a_{\max}$ &~~~~~& Core & $N$ & $a_{\min}$ & $a_{\max}$ \\
\cline{1-4}
\cline{6-9}
\endhead
\cline{1-4}
\cline{6-9}
\multicolumn{9}{r}{\textit{Continued on next page}}\\
\endfoot
\endlastfoot
9900 & 203 & 2.141 & 2.144 & & 1145 & 539 & 2.424 & 2.425 \\
3972 & 180 & 2.163 & 2.164 & & 6 & 330 & 2.4252 & 2.4258 \\
2770 & 192 & 2.1695 & 2.1705 & & 1108 & 382 & 2.427 & 2.4277 \\
1468 & 381 & 2.1815 & 2.183 & & 585 & 482 & 2.4293 & 2.4302 \\
512 & 253 & 2.1885 & 2.1895 & & 112 & 543 & 2.4338 & 2.4348 \\
1733 & 254 & 2.193 & 2.194 & & 79 & 273 & 2.4441 & 2.4447 \\
270 & 279 & 2.198 & 2.199 & & 4088 & 190 & 2.4449 & 2.4453 \\
8 & 280 & 2.201 & 2.202 & & 2026 & 221 & 2.4455 & 2.446 \\
43 & 168 & 2.2032 & 2.2037 & & 138 & 456 & 2.4472 & 2.4482 \\
1219 & 1011 & 2.2105 & 2.213 & & 13698 & 217 & 2.4485 & 2.449 \\
443 & 311 & 2.215 & 2.216 & & 2898 & 267 & 2.556 & 2.5565 \\
1123 & 399 & 2.2245 & 2.2255 & & 1658 & 510 & 2.5595 & 2.5603 \\
422 & 578 & 2.228 & 2.2295 & & 429 & 640 & 2.607 & 2.608 \\
937 & 388 & 2.231 & 2.232 & & 70 & 811 & 2.6145 & 2.6155 \\
685 & 511 & 2.2355 & 2.2365 & & 53 & 812 & 2.618 & 2.619 \\
1523 & 411 & 2.242 & 2.243 & & 792 & 1069 & 2.6225 & 2.6235 \\
2037 & 239 &  2.2455 & 2.246 & & 615 & 616 & 2.6305 & 2.6315  \\
822 & 1524 & 2.254 & 2.257 & & 476 & 2505 & 2.649 & 2.653 \\
3982 & 486 & 2.2585 & 2.2595 & & 102 & 638 & 2.661 & 2.662 \\
1899 & 493 & 2.2645 & 2.2655 & & 64 & 417 & 2.6811 & 2.6818 \\
1078 & 503 & 2.269 & 2.27 & & 166 & 596 & 2.6855 & 2.6865 \\
3841 & 510 & 2.2735 & 2.2745 & & 868 & 1631 & 2.704 & 2.706 \\
5764 & 288 & 2.276 & 2.2765 & & 1904 & 396 & 2.7433 & 2.744 \\
548 & 624 & 2.282 & 2.2825 & & 934 & 367 & 2.7478 & 2.7485 \\
2013 & 434 & 2.2893 & 2.29 & & 485 & 1567 & 2.751 & 2.753 \\
1419 & 470 & 2.2925 & 2.2932 & & 356 & 454 & 2.7565 & 2.7573  \\
45153 & 320 & 2.299 & 2.2995 & & 143 & 562 & 2.761 & 2.762  \\
4262 & 769 & 2.3015 & 2.3025 & & 446 & 441 & 2.787 & 2.7878 \\
6189 & 676 & 2.3045 & 2.3055 & & 1092 & 1306 & 2.901 & 2.909  \\
1982 & 403 & 2.3095 & 2.31 & & 22 & 226 & 2.909 & 2.91 \\
1959 & 749 & 2.316 & 2.317 & & 677 & 241 & 2.9555 & 2.9575 \\
4408 & 639 & 2.322 & 2.323 & & 447 & 506 & 2.9855 & 2.9863 \\
1083 & 680 & 2.3277 & 2.3285 & & 117 & 346 & 2.991 & 2.992  \\
2762 & 551 & 2.3305 & 2.3312 & & 221 & 381 & 3.012 & 3.013 \\
1664 & 385 & 2.3325 & 2.333 & & 478 & 539 & 3.0155 & 3.017 \\
290 & 666 & 2.3368 & 2.3375 & & 592 & 3431 & 3.0208 & 3.03  \\
9963 & 729 & 2.341 & 2.342 & & 1488 & 334 & 3.0385 & 3.0391 \\
1367 & 794 & 2.344 & 2.345 & & 4410 & 617 & 3.054 & 3.0552 \\
27 & 736 & 2.347 & 2.348 & & 368 & 578 & 3.067 & 3.068 \\
3895 & 597 & 2.3505 & 2.3512 & & 202 & 3354 & 3.074 & 3.078 \\
4857 & 527 & 2.3558 & 2.3565 & & 2395 & 460 & 3.0795 & 3.082 \\
1646 & 773 & 2.36 & 2.361 & & 1684 & 343 & 3.0908 & 3.0913 \\
916 & 723 & 2.364 & 2.365 & & 86 & 1910 & 3.105 & 3.1072 \\
163 & 753 & 2.367 & 2.368 & & 196 & 648 & 3.1135 & 3.1145 \\
1573 & 466 & 2.3703 & 2.371 & & 382 & 743 & 3.122 & 3.1232 \\
584 & 719 & 2.373 & 2.374 & & 375 & 1531 & 3.128 & 3.1302  \\
249 & 437 & 2.3772 & 2.3778 & & 10 & 770 & 3.141 & 3.1422  \\
4904 & 780 & 2.388 & 2.389 & & 209 & 677 & 3.147 & 3.1485 \\
1591 & 874 & 2.3908 & 2.392 & & 2494 & 583 & 3.16 & 3.161 \\
1077 & 565 & 2.3921 & 2.3929 & & 1023 & 1105 & 3.167 & 3.169 \\
463 & 1416 & 2.397 & 2.3983 & & 511 & 2761 & 3.173 & 3.1752 \\
304 & 1363 & 2.403 & 2.405 & & 778 & 500 & 3.18 & 3.181 \\
4132 & 320 & 2.407 & 2.4075 & & 530 & 1048 & 3.2065 & 3.2088 \\
6334 & 331 & 2.4075 & 2.408 & & 1362 & 316 & 3.273 & 3.277 \\
182 & 2724 & 2.4178 & 2.4192 & & 190 & 1986 & 3.956 & 3.9687 \\
\cline{1-4}
\cline{6-9}
\end{longtable}

\section{Bivariate distributions and 3D analysis via 2D projections}
\label{sec_2D}
The 2D wavelet analysis appears more complicated, because the 2D geometry is considerably
more diverse than the 1D one. Also, the 2D case is more computationally demanding. In
\citep{BalRodShai19} the 2D wavelet analysis algorithm is presented, based on
the optimised radially-symmetric (isotropic) wavelets 2DOPT$_{1,2}$. These two wavelets are
almost identical, and here we use the 2DOPT$_2$ version which we refer to as just 2DOPT for
simplicity.

Regardless to the complications, the 2D analysis appears analogous to 1D one in many
aspects. However, because of the isotropic restriction on the wavelet shape, this algorithm
can be only applied to physically comparable (summable) parameters, and targets mainly
patterns that have similar size in the both directions.

The 1D analysis above was focused on the following orbital parameters: eccentricity $e$,
inclination $i$, and semimajor axis $a$. We have not constructed a 3D algorithm yet to
process this 3D space $(a,e,i)$ in a self-consistent manner, but we can consider three
independent 2D subspaces: $(a,e)$, $(a,i)$, and $(e,i)$. We may consider a 2D density in
each of these planes and investigate it using our 2D algorithm.

We adopt the following system of comparable parameters: $(\log a, e, \sin i)$. Here, $\log
a$ appears instead of $a$ because the differences like $\Delta \log a \simeq \Delta a/a$
appear adimensional, as well as the differences $\Delta e$ or $\Delta \sin i$. Hence, we
can legally compare various small ranges in terms of $\log a$ with ranges for $e$ and $\sin
i$ (hence, all three wavelet scales appear dimensionless). Concerning the physical
comparability of $e$ and $\sin i$, it follows because these (or equivalent) parameters
often play equal roles in various dynamical equations; this is highlighted by e.g. the
Lidov-Kozai mechanism where these parameters can ``flow'' one into another through the
conservation of the quantity $\sqrt{1-e^2} \cos i$, so $e$ can be exchanged with $\sin i$
\citep[chap.~7]{MurrayDermott}.

\begin{figure}
\begin{tabular}{@{}c@{}c@{}}
\includegraphics[width=0.49\linewidth]{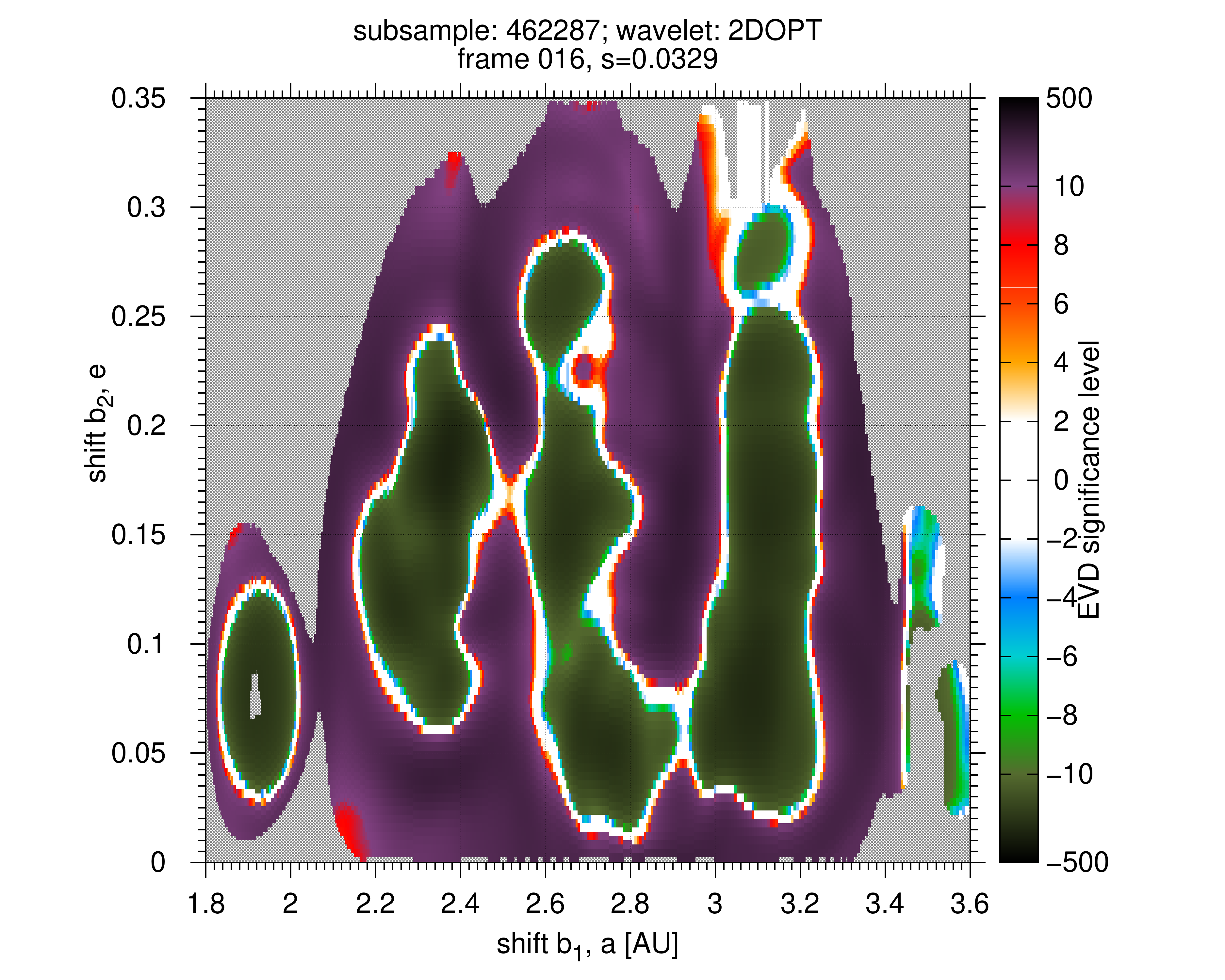} & \includegraphics[width=0.49\linewidth]{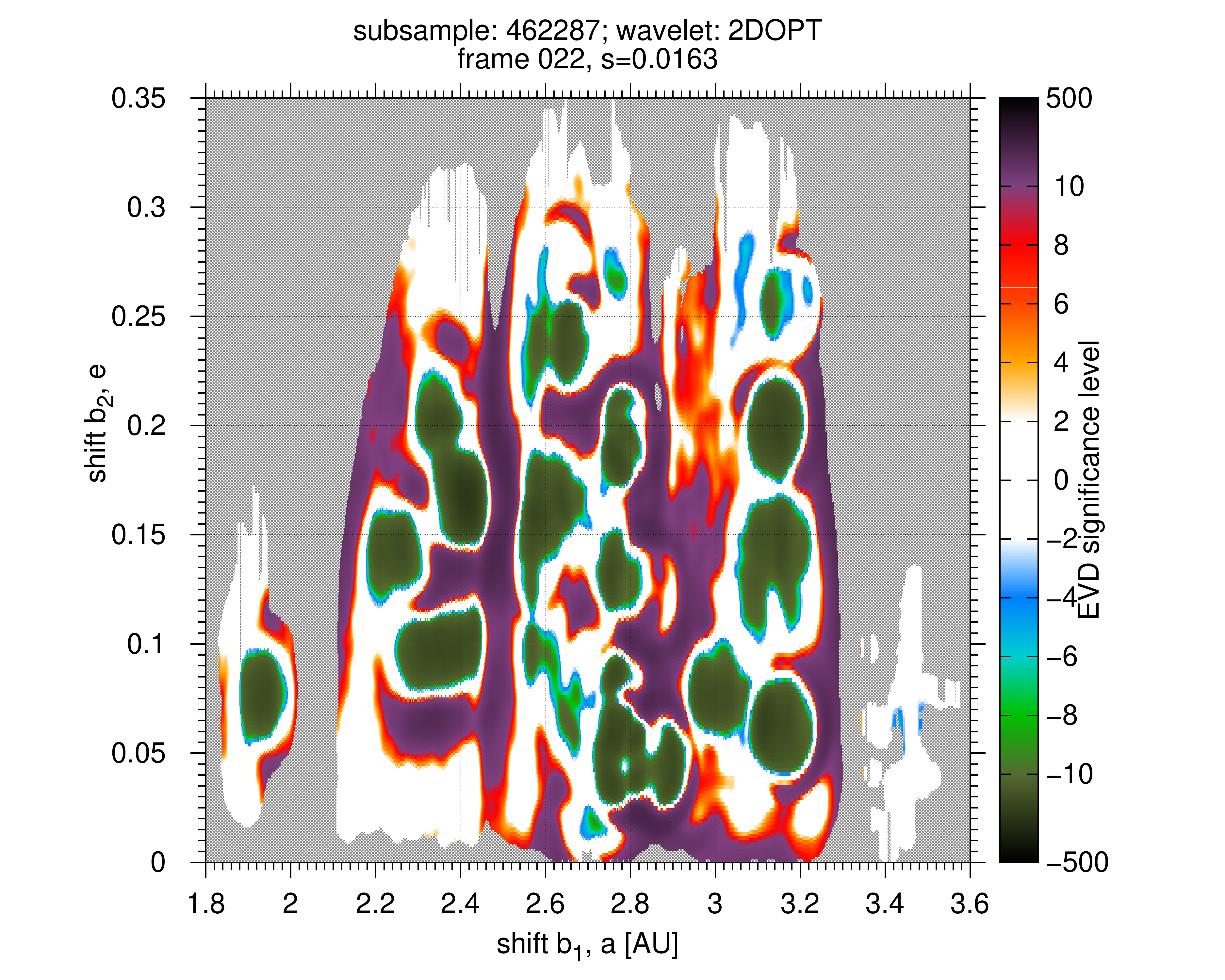} \\
\includegraphics[width=0.49\linewidth]{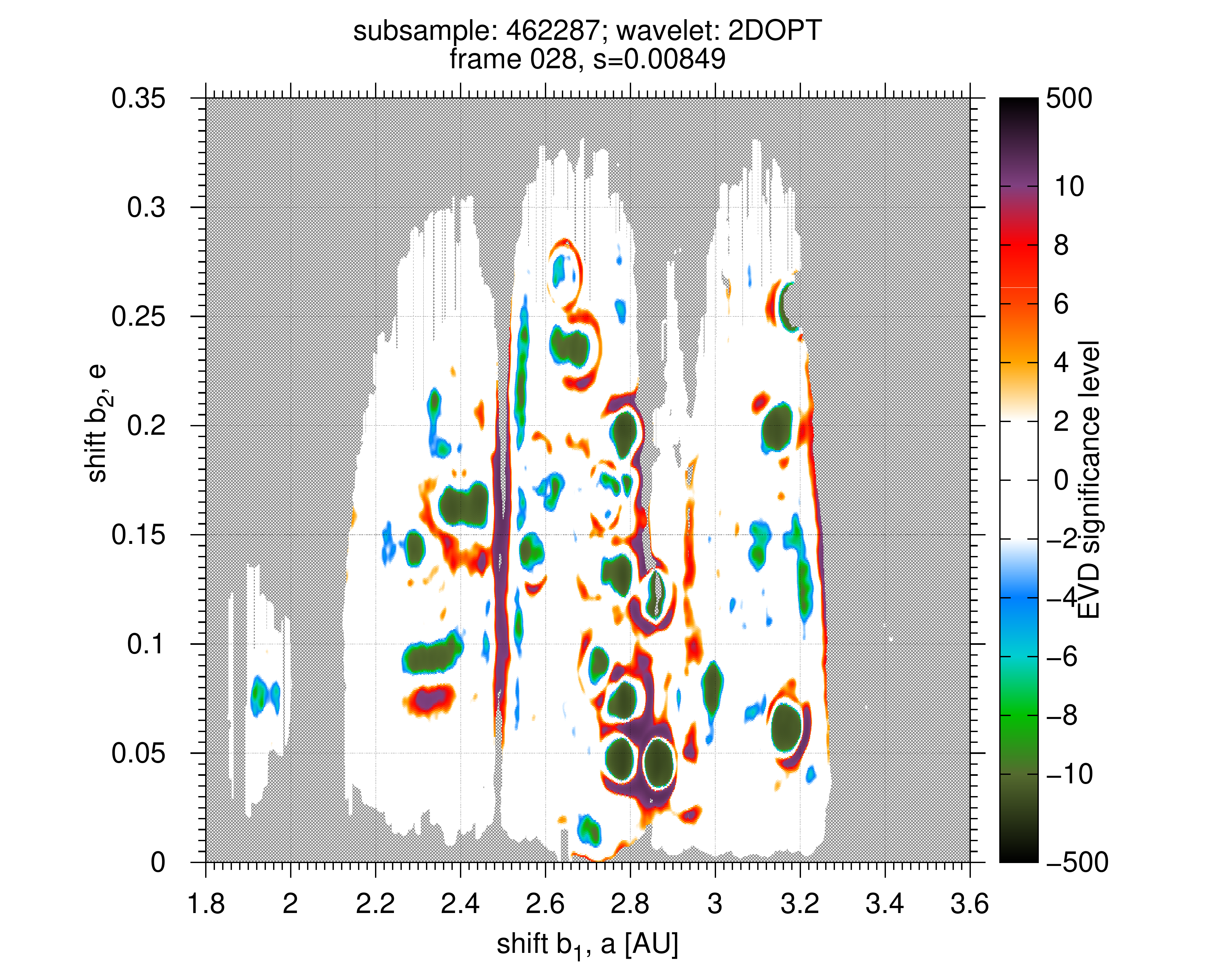} & \includegraphics[width=0.49\linewidth]{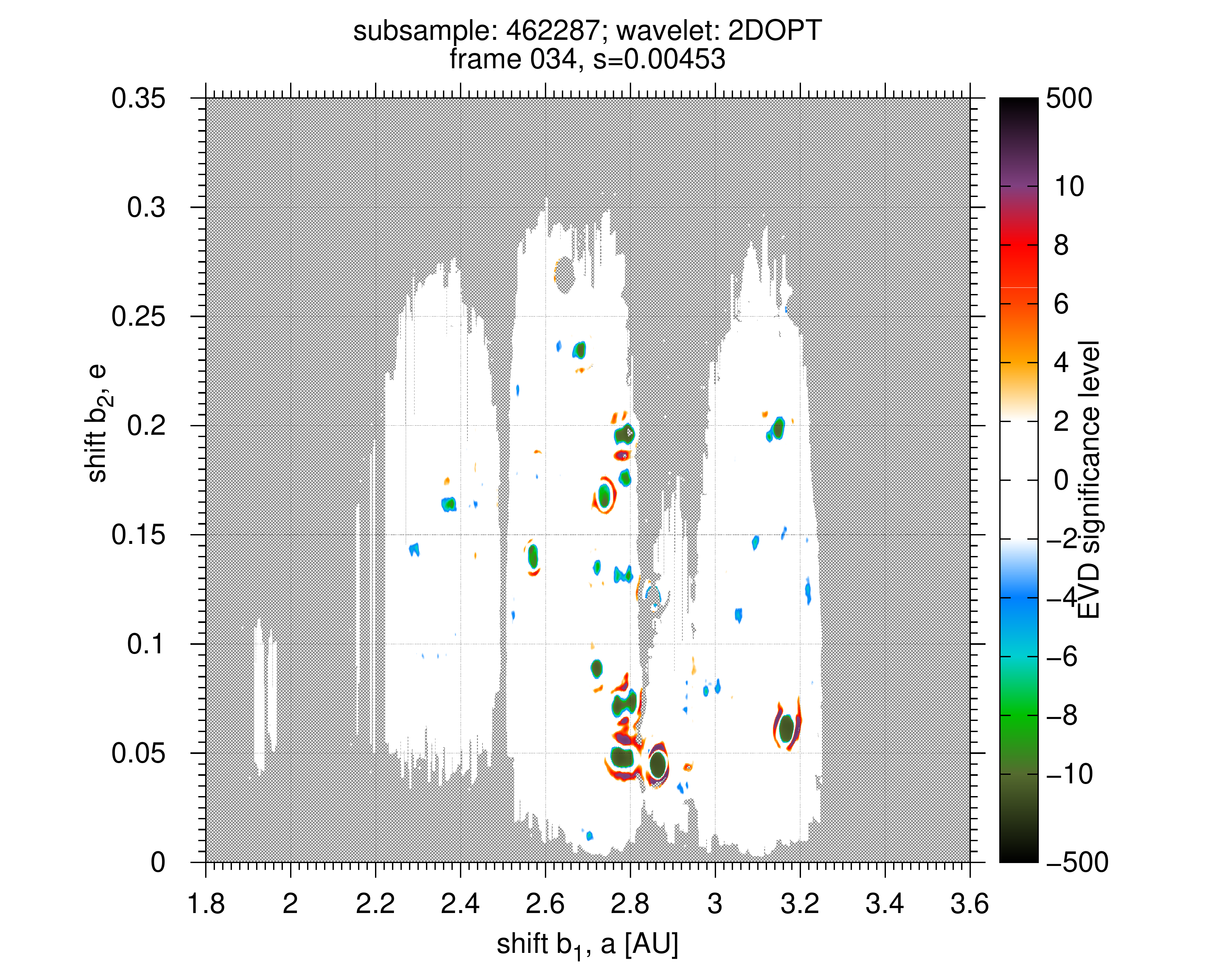} \\
\includegraphics[width=0.49\linewidth]{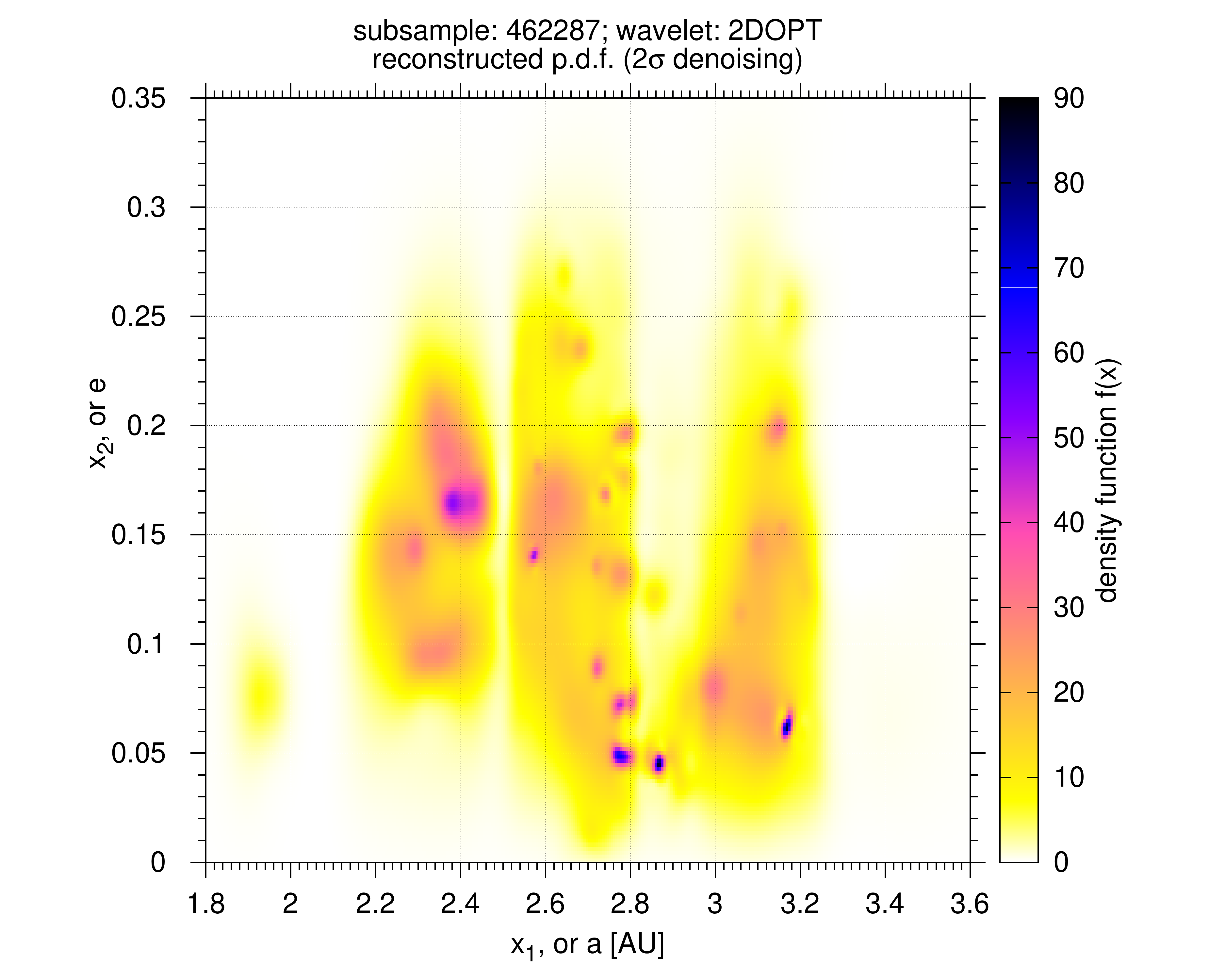}     & \includegraphics[width=0.49\linewidth]{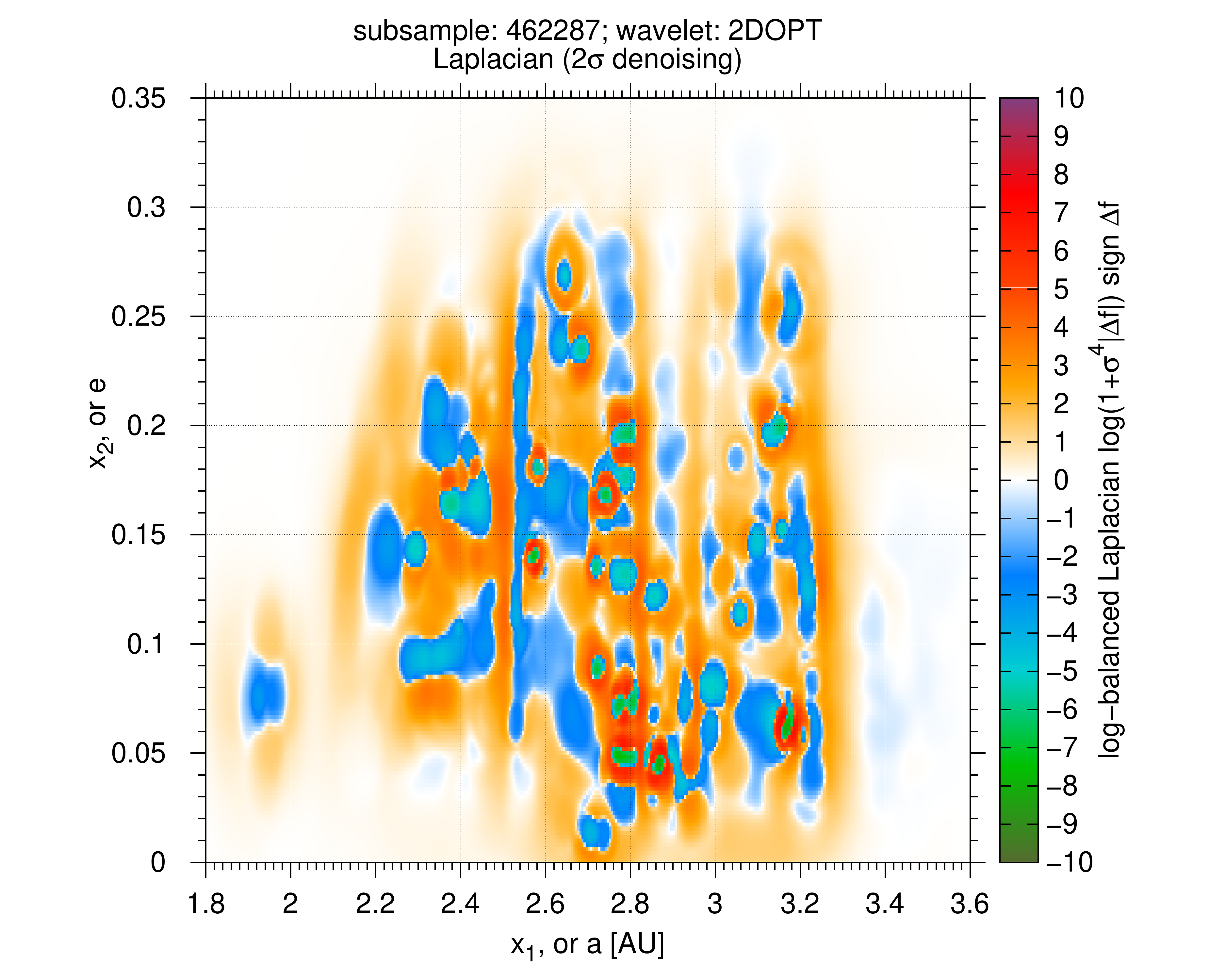} \\
\end{tabular}
\caption{\label{2d-ae}
Top quadruplet: $4$ frames of the 2D wavelet transform for the $(\log A, e)$ bivariate
distribution (proper elements). The animation containing the full CWT evolution is also
attached as online-only material. Bottom pair: p.d.f. model reconstructed by matching
pursuit iterations and its Laplacian.}
\end{figure}

For example, let us consider the $(\log a, e)$ pair (Fig.~\ref{2d-ae}). Notice that the
wavelet transform is a function of three variables now, so we plot several frames
corresponding to difference scales. Each such frame is plotted as a significance map (as in
the 1D analysis).

However, investigating the 2D wavelet transform directly does not appear very easy, since
we should treat multiple resolution levels simultaneously. The reconstructed p.d.f. model
would be more helpful here, because it joins all resolution levels into the same plot,
simultaneously keeping only the significant detected structures. However, in practice the
p.d.f. graphs appeared too much diffuse and inconclusive, because they do not highlight
subtle asteroid families even if they are statistically significant. Such a subtle cluster
would appear almost indistinguishable over the large-scale background, because it changes
the background level only very slightly. We found that this issue can be solved by
considering the Laplacian of the p.d.f. model rather than this model itself. This is
justified by the known property that the CWT represents a smoothed Laplacian
\citep{BalRodShai19}, so in fact our wavelet analysis deals with the p.d.f.
Laplacian rather than p.d.f. itself. The Laplacian can be easily computed by applying the
CWT with a small scale (smaller than scales of all the detected structures).

\begin{figure}
\includegraphics[width=0.49\linewidth]{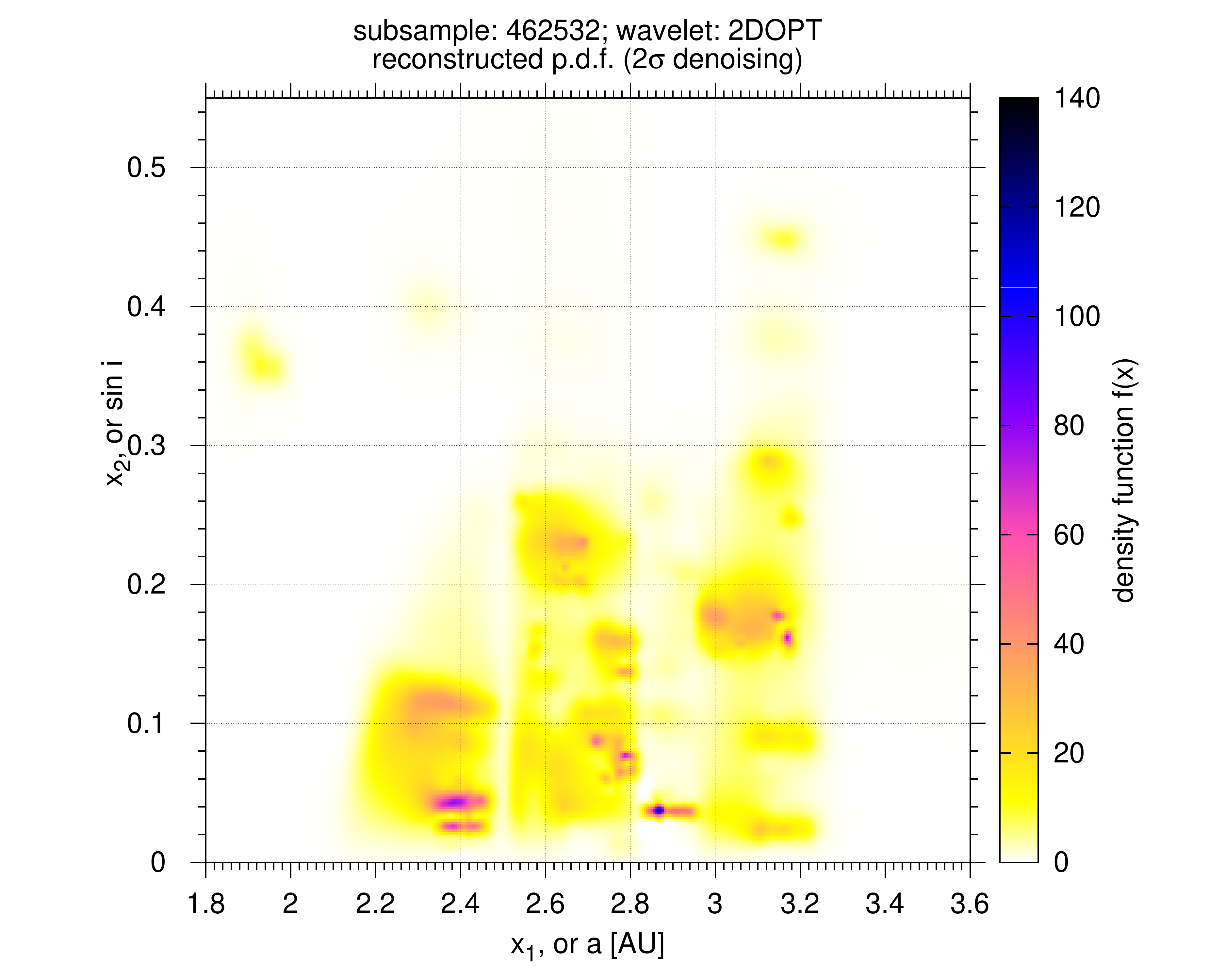}
\includegraphics[width=0.49\linewidth]{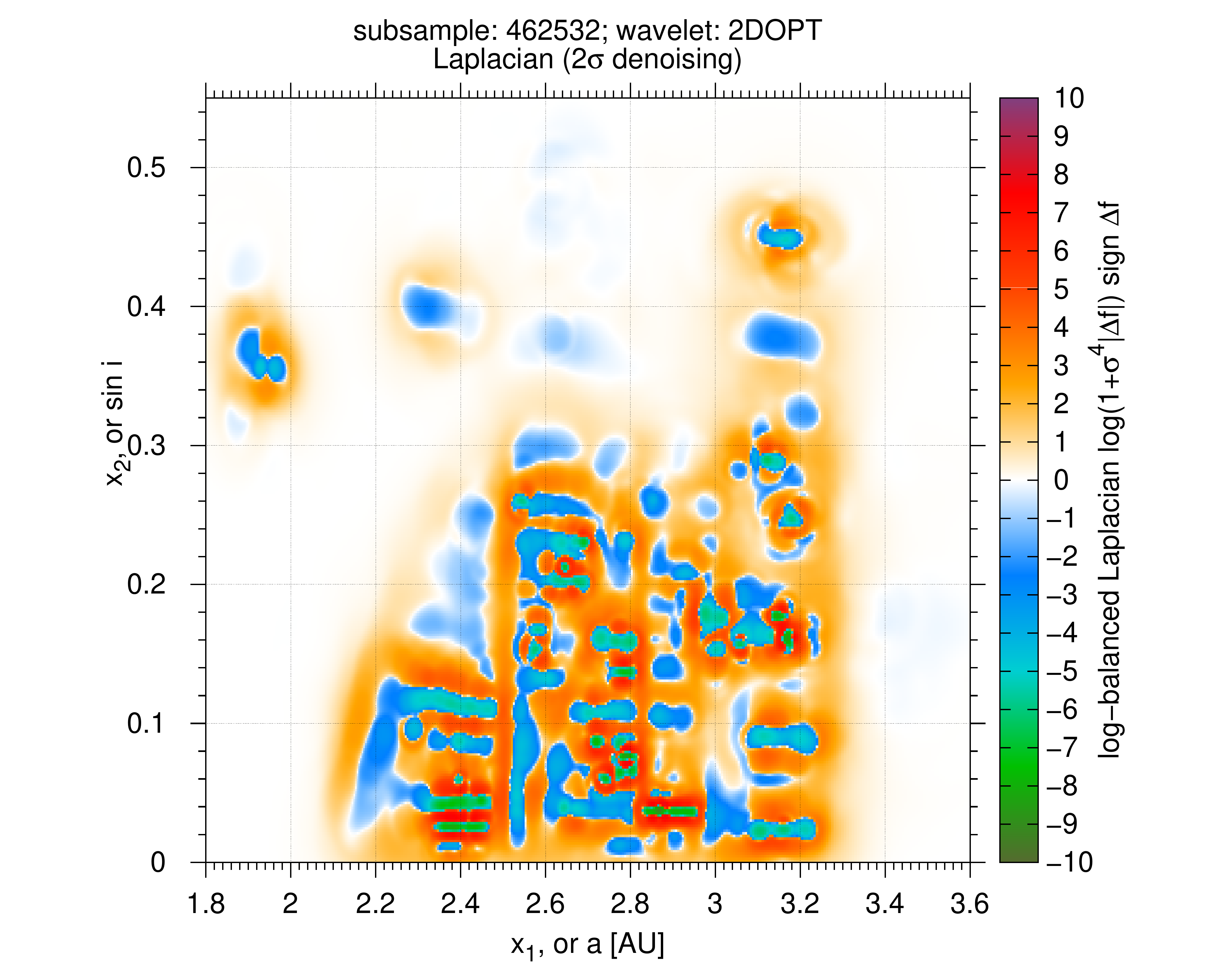}
\caption{\label{2d-ai}
The reconstructed p.d.f. model for the $(\log a, \sin i)$ bivariate distribution
and its Laplacian.}
\end{figure}

\begin{figure}
\includegraphics[width=0.49\linewidth]{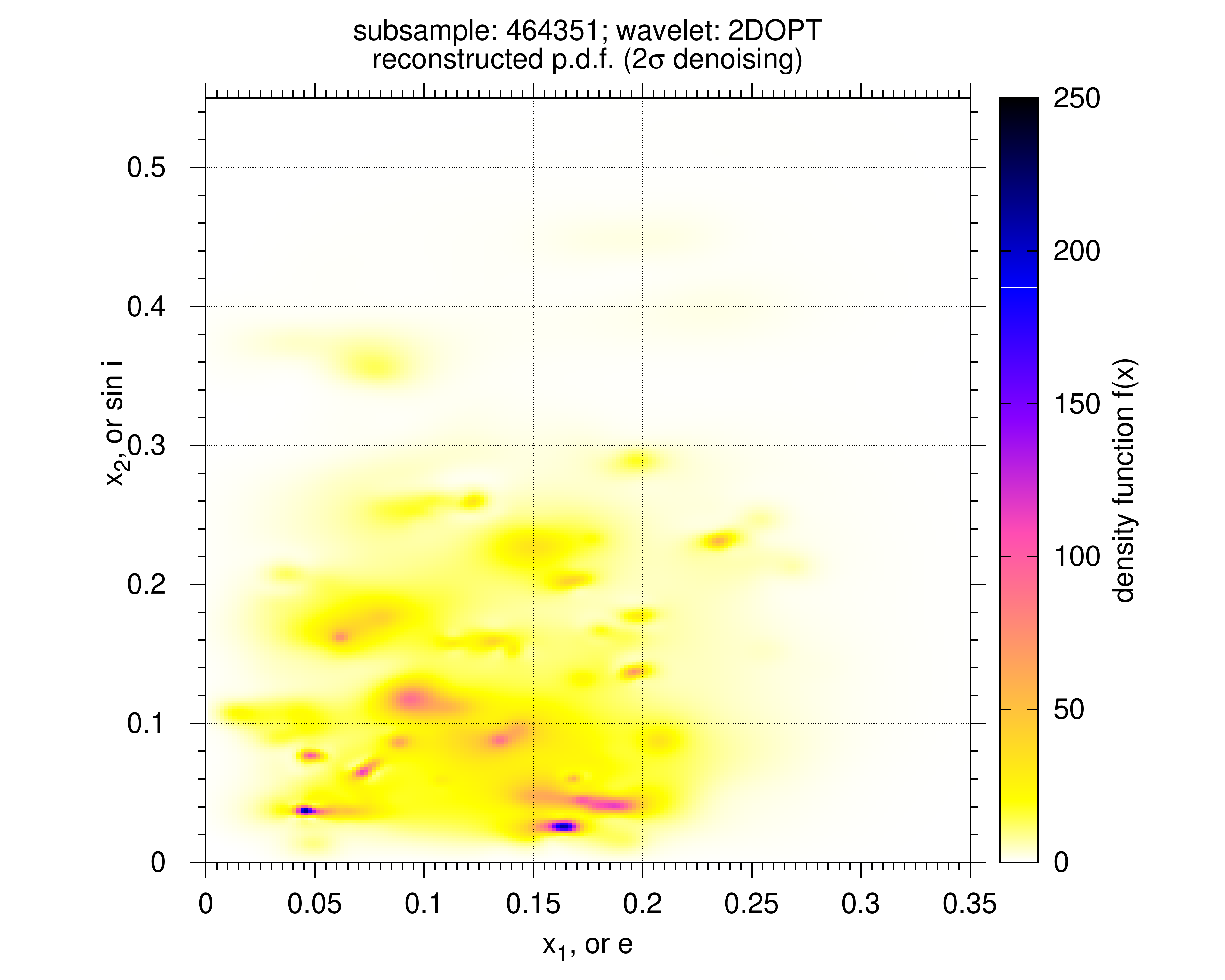}
\includegraphics[width=0.49\linewidth]{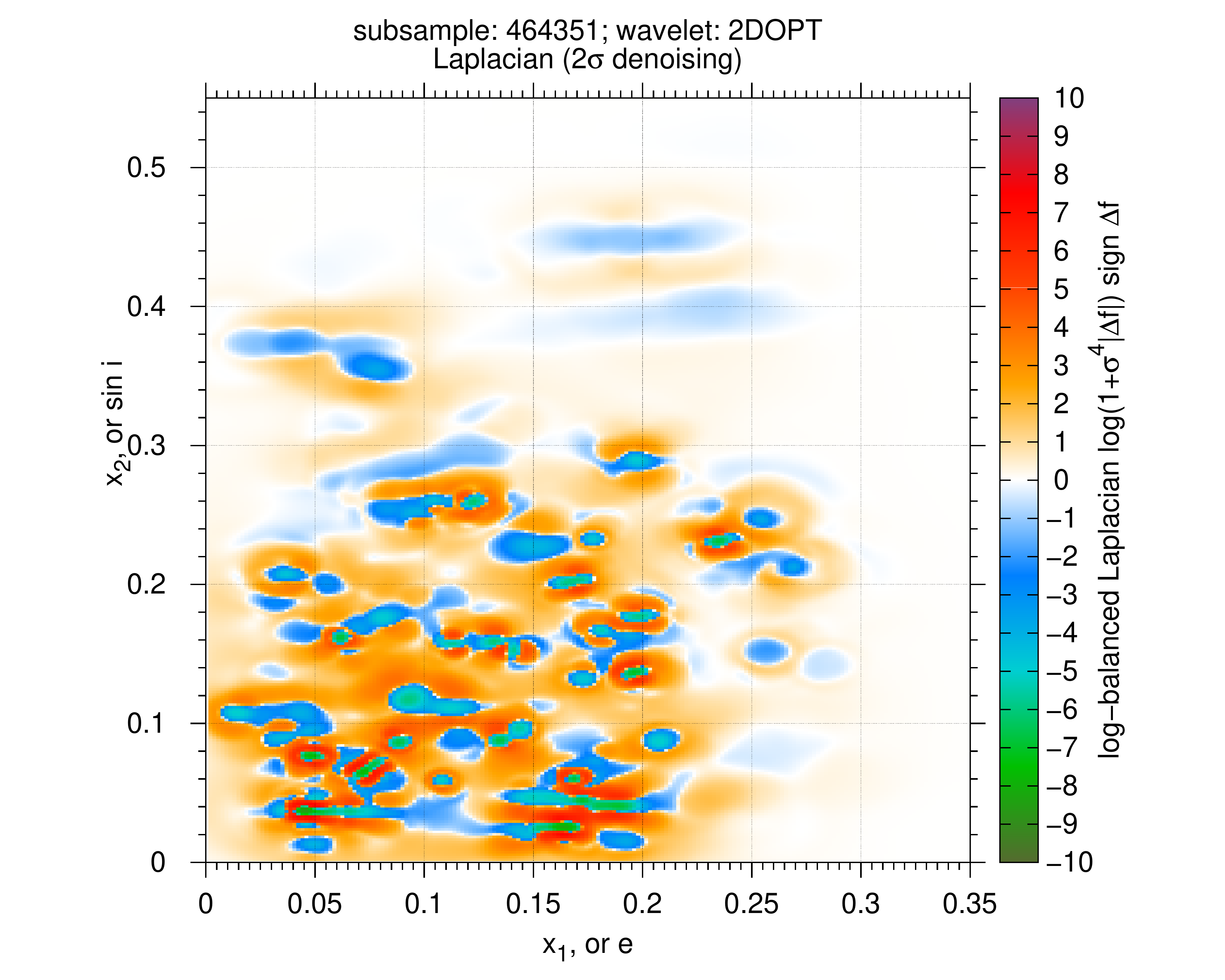}
\caption{\label{2d-ei}
The reconstructed p.d.f. model for the $(e, \sin i)$ bivariate distribution
and its Laplacian.}
\end{figure}

As we can see, the Laplacian appears very helpful to visually spot even very subtle
families in any of the bivariate distributions that we considered (Fig.~\ref{2d-ae},
\ref{2d-ai}, \ref{2d-ei}). To further highlight the color contrast, we plot here a
logarithmically-modified quantity $\log (1+\sigma^4 |\Delta f|) \sign \Delta f$, where
$\sigma^2=\sigma_1^2+\sigma_2^2$ is the cumulative variance of the two random variables.

We can see that boundaries of a cluster can be determined as boundaries of an isolated
domain (``spot''). Notice that it is important to pay attention to an opposite-sign ring
around each spot. If it is present then we have a local convexity (negative Laplacian)
surrounded by a concavity ring (positive Laplacian). Such a structure can be interpreted
as an isolated cluster. However, if this ring is not present (not significant) then we
cannot claim that such a geometric structure is a family, because it is not separated from
the background. We adopt this rule as a basic formalized definition of a ``cluster'' in
this work. This treatment is justified in more details in a separate work devoted to the
stellar population analysis \citep{BalShaiVes20}.

A more difficult question appears if some hints of a ring are present, but the ring is
incomplete, or if there are two partly merged 2D spots not separated from each other by a
zone of positive Laplacian. This typically appears in case of overlapping families. Since
such families can often be distinguished with the help of the third parameter (the one not
involved in the given 2D plot), we investigate each such case individually.

We try to understand the 3D p.d.f. via its 2D projections, so the overlapping effect
becomes very important. For each potential family (or a group of overlapping families) in
each of the three 2D diagrams we cut out a rectangular box in the corresponding 2D plane
and consider the subsample containing only asteroids within this box. For each such
subsample we performed a 1D wavelet analysis of the third parameter and constructed the
corresponding 1D p.d.f. containing only statistically significant patterns. Such 1D
distributions suggest useful hints allowing to resolve various ambiguities concerning
family overlapping. For example, if this distribution is unimodal then the given candidate
family is homogeneous (no overlapping). If there are additional modes then the apparent 2D
family actually contains two overlapping families corresponding to different values of the
third parameter, and so on. These hints can be additionally verified by looking at the
other two 2D planes.

Of course, there are more difficult cases that cannot be resolved unambiguously based on
just the 2D projections. This may occur in case of a partial overlap of multiple families
in all 2D diagrams and other nuisance effects \citep{BalRodShai19}. Nevertheless, we found
$44$ asteroid families that can be resolved clearly. They are listed in
Table~\ref{residual}. The families were cross-identified with the known AstDys ones
(Table~\ref{AstDysFamilies}) by comparing their boundaries. We find that almost every of
our wavelet-detected family has and HCM-based counterpart, but the wavelet-derived ranges
are systematically more narrow. Notice that the boundaries of a family can be rather
diffuse and thus their exact position is largely a matter of convention. Our convention is
to define the boundary based on zero Laplacian (or zero second derivative in the 1D
distribution). Our results suggests that this convention leads to a more restrictive
boundaries than from HCM, this is the same effect as in \citep{BalShaiVes20}.

\begin{longtable}{|c|c|cc|cc|cc|}
\caption{\label{residual} Asteroid families detected on the basis of 2D wavelet analysis.}\\
\hline
No   & HCM core &   $a_{\min}$   &   $a_{\max}$      &   $e_{\min}$     &   $e_{\max}$     &  $\sin i_{\min}$ & $\sin i_{\max}$   \\
\hline
\endfirsthead
\multicolumn{7}{c}{\tablename\ \thetable\ -- \textit{continued}}\\
\hline
No   & HCM core &   $a_{\min}$   &   $a_{\max}$      &   $e_{\min}$     &   $e_{\max}$     &  $\sin i_{\min}$ & $\sin i_{\max}$   \\
\hline
\endhead
\hline \multicolumn{7}{r}{\textit{Continued on next page}}\\
\endfoot
\endlastfoot
\hline
W1   &  \hyperlink{h434}{434}        &    1.87        &      1.99         &    0.057         &    0.094         &   0.345          &    0.380 \\
W2   &  \hyperlink{h2076}{2076}      &    2.265       &      2.31         &    0.138         &    0.150         &   0.088          &    0.102 \\
W3   &  \hyperlink{h4}{4}            &    2.25        &      2.48         &    0.083         &    0.129         &   0.105          &    0.126 \\
W4   &  \hyperlink{h163}{163}        &    2.32        &      2.37         &    0.199         &    0.217         &   0.08           &    0.096 \\
W5   &  27 (FIN410)                  &    2.34        &      2.4          &    0.179         &    0.201         &   0.008          &    0.014 \\
W6   &  \hyperlink{h20}{20}          &    2.345       &      2.465        &    0.149         &    0.171         &   0.022          &    0.029 \\
W7   &  \hyperlink{h5026}{5026}      &    2.37        &      2.41         &    0.199         &    0.217         &   0.079          &    0.093 \\
W8   &  \hyperlink{h302}{302}        &    2.385       &      2.405        &    0.103         &    0.112         &   0.055          &    0.063 \\
W9   &  \hyperlink{h1658}{1658}      &    2.53        &      2.645        &    0.164         &    0.179         &   0.125          &    0.138 \\
W10  &  \hyperlink{h3815}{3815}      &    2.555       &      2.585        &    0.136         &    0.144         &   0.143          &    0.161 \\
W11  &  \hyperlink{h606}{606}        &    2.57        &      2.595        &    0.175         &    0.184         &   0.162          &    0.171 \\
W12  &  \hyperlink{h3}{3}            &    2.6         &      2.70         &    0.227         &    0.245         &   0.226          &    0.237 \\
W13  &  \hyperlink{h145}{145}        &    2.6         &      2.705        &    0.155         &    0.178         &   0.196          &    0.208 \\
W14  &  \hyperlink{h1547}{1547}      &    2.635       &      2.655        &    0.261         &    0.276         &   0.209          &    0.215 \\
W15  &  \hyperlink{h808}{808}        &    2.705       &      2.735        &    0.13          &    0.139         &   0.082          &    0.092 \\
W16  &  \hyperlink{h3827}{3827}      &    2.705       &      2.74         &    0.083         &    0.094         &   0.082          &    0.092 \\
W17  &  173 (FIN522)                 &    2.715       &      2.745        &    0.171         &    0.183         &   0.227          &    0.238 \\
W18  &  \hyperlink{h396}{396}        &    2.725       &      2.75         &    0.164         &    0.172         &   0.056          &    0.064 \\
W19  &  \hyperlink{h668}{668}        &    2.74        &      2.81         &    0.19          &    0.202         &   0.131          &    0.141 \\
W20  &  \hyperlink{h93}{93}          &    2.745       &      2.815        &    0.122         &    0.14          &   0.151          &    0.17  \\
W21  &  \hyperlink{h847}{847}        &    2.75        &      2.79         &    0.067         &    0.076         &   0.06           &    0.069 \\
W22  &  \hyperlink{h808}{808}        &    2.75        &      2.81         &    0.128         &    0.139         &   0.082          &    0.093 \\
W23  &  \hyperlink{h1128}{1128}      &    2.75        &      2.815        &    0.044         &    0.053         &   0.006          &    0.018 \\
W24  &  2353                         &    2.76        &      2.81         &    0.087         &    0.103         &   0.08           &    0.093 \\
W25  &  \hyperlink{h18466}{18466}    &    2.76        &      2.81         &    0.171         &    0.181         &   0.227          &    0.238 \\
W26  &  \hyperlink{h1726}{1726}      &    2.77        &      2.815        &    0.044         &    0.053         &   0.073          &    0.079 \\
W27  &  \hyperlink{h847}{847}        &    2.79        &      2.815        &    0.067         &    0.082         &   0.06           &    0.076 \\
W28  &  \hyperlink{h158}{158}        &    2.83        &      2.85         &    0.043         &    0.055         &   0.033          &    0.04  \\
W29  &  \hyperlink{h293}{293}        &    2.83        &      2.89         &    0.116         &    0.128         &   0.254          &    0.265 \\
W30  &  \hyperlink{h158}{158}        &    2.84        &      2.865        &    0.063         &    0.072         &   0.033          &    0.04  \\
W31  &  \hyperlink{h158}{158}        &    2.85        &      2.88         &    0.04          &    0.049         &   0.033          &    0.04  \\
W32  &  \hyperlink{h16286}{16286}    &    2.85        &      2.88         &    0.04          &    0.049         &   0.093          &    0.115 \\
W33  &  \hyperlink{h845}{845}        &    2.89        &      2.96         &    0.026         &    0.047         &   0.202          &    0.214 \\
W34  &  \hyperlink{h158}{158}        &    2.91        &      2.945        &    0.062         &    0.089         &   0.032          &    0.04  \\
W35  &  \hyperlink{h221}{221}        &    2.96        &      3.025        &    0.070         &    0.093         &   0.163          &    0.187 \\
W36  &  \hyperlink{h179}{179}        &    2.975       &      3.01         &    0.061         &    0.07          &    0.149         &    0.153 \\
W37  &  \hyperlink{h96}{96}          &    3.03        &      3.065        &    0.179         &    0.191         &   0.275          &    0.288 \\
W38  &  \hyperlink{h283}{283}        &    3.04        &      3.07         &    0.107         &    0.120         &   0.153          &    0.16  \\
W39  &  \hyperlink{h24}{24}          &    3.07        &      3.12         &    0.138         &    0.156         &   0.016          &    0.029 \\
W40  &  \hyperlink{h1040}{1040}      &    3.105       &      3.165        &    0.189         &    0.207         &   0.280          &    0.295 \\
W41  &  \hyperlink{h3330}{3330}      &    3.13        &      3.17         &    0.189         &    0.207         &   0.173          &    0.181 \\
W42  &  \hyperlink{h24}{24}          &    3.14        &      3.17         &    0.147         &    0.157         &   0.017          &    0.028 \\
W43  &  \hyperlink{h778}{778}        &    3.145       &      3.205        &    0.246         &    0.264         &   0.240          &    0.255 \\
W44  &  \hyperlink{h490}{490}        &    3.155       &      3.18         &  0.057           &    0.066         &   0.157          &    0.167 \\
\hline
\end{longtable}

We notice that our wavelet analysis detected three asteroid families not mentioned in
AstDys (W5, W17, W24). After a closer look, it appeared that W5 and W17 are the 27~Euterpe
and 173~Ino families mentioned by \citet{Nesvorny15} as FIN410 and FIN522. However, there
is no more details about these families, and they are not included in AstDys. The
corresponding asteroids are labelled in AstDys as not involved in any family. Therefore, we
see some controversy in the literature concerning these two families, and our analysis
resolves it positively. The third family W24 has the smallest-number asteroid 2353, and
likely appears unknown.

Simultaneously, there are many HCM-based families not detected by wavelets. In some part,
this can be explained by the overlapping effect which disabled unambiguous detection of
some families by wavelets. Likely, the full 3D wavelet analysis would detect more families,
but we currently do not have a working 3D extension of our wavelet analysis pipeline (this
needs substantial additional theory work and computing optimisations). However, the
overlapping does not explain all such occurrences well. Many of the HCM-only families just
do not reveal themselves in our wavelet analysis, that is they appear statistically
insignificant in our approach. From the other side, some of them may appear more
significant in the full 3D analysis. But at the current stage such families are possibly
more doubtful and require additional investigation that falls out of the scope of the
present paper.

Also, we notice that some HCM-detected families may reveal a complicated structure. In our
analysis they are split into multiple subfamilies (up to $4$, like the Koronis
family\footnote{Among them, the family W31 might refer to the Karin group
\citep{Nesvorny02}.}). In some part this may indicate that our wavelet analysis tends to
generate some crowding effect, contrary to the HCM chaining.

Concerning the resonant asteroid families, we did not detect them in the 2D analysis,
likely because they should reveal themselves as extremely elongated thin patterns. Notice
that our 2D analysis is based on isotropic radially-symmetric wavelets, so it is expectedly
insensitive to such disproportional structures.

\section{Conclusions and discussion}
Our main conclusion is that statistical wavelet analysis appears as a useful alternative
tool allowing to independently verify the HCM results. Let us now review their main
differences and outline sevelar prospects to advance futher.

\begin{enumerate}
\item The wavelets generate a crowding effect opposite to the HCM chaining effect (as
expected). This results in a fragmentation of large statistical clusters into smaller
subgroups. One reason for such a difference is that we use radially symmetric wavelets that
naturally tends to decompose an elongated structure into a sequence of more or less oval
ones.

\item In the framework of the wavelet analysis, the balance between the crowding and
chaining effects can be controlled through the use of non-radially-symmetric elliptically
distorted wavelets. Currently such wavelets are not used at all, but it is possible to
include them by replacing a single scale parameter $s$ with a general scale matrix
in~(\ref{cwt}). In such a case elongated and radially symmetric wavelets can be combined
together using a tunable weight function (to appear in~(\ref{icwt})). Increasing the role
of elliptic wavelets would bias the method to have more chaining effect. However, the use
of elliptic wavelets implies a jump of dimensionality and hence the need to rework the
entire computing approach (see below).

\item Both the methods, wavelets and HCM, involve some dependence on various assumptions.
While HCM may depend on the metric used, the wavelet analysis depends on the wavelet shape.
Moreover, selecting different wavelets we may control the underlying metric. For example,
radially symmetric wavelets imply the use of a local $L_2$ metric in the space of the
variables that we analyse. In our wavelet algorithm the radial symmetry is also a just a
particular prior assumption, but beyond this restriction the wavelet radial function
$\psi(t)$ was derived from certain optimality criteria to minimize the noise (and to
increase the S/N ratio for possible patterns). In view of this, it might be an interesting
idea trying to find some optimal metric for HCM.

\item Our wavelet analysis is currently limited by two dimensions, while both the asteroid
families search presented here and stellar population analysis presented in
\citep{BalShaiVes20} assume at least 3D spaces. The generalization of the wavelet analysis
and the associated tools to $\mathbb R^n$ with $n>2$ is not difficult mathematically, but
it infers significant increase of computational issues. The computing approaches of our
algorithm should be reworked qualitatively then. The main issues are the efficient discrete
coverage of the shift-scale space for~\ref{cwt} and numeric integration of~(\ref{icwt}).
Currently this is achieved through a regular rectangular grid, but results in exponential
dependence of the required resources on the dimensionality.

\item We found considerably smaller number of asteroid families than known from the HCM
method. It looks as if many HCM families have too low statistical significance in our
analysis and look like just noise. However, we are unsure about this conclusion because
similar effect can appear by other reasons. In some part it can be explained by law
dimensionality of our analysis (e.g. a statistical group can appear more dense in some
additional variable that we did not consider here). In some part this appeared due to
overlapping effect (we could not disentangle all 3D asteroid families based on 2D
projections). So this issue requires further investigation.

\item In addition to all said above, our analysis allowed to reveal some new families not
detected with HCM, to confirm possibly controversial families, and to reveal internal
structure in big HCM families.

\item The wavelet analysis is not a cluster detection tool in the strict meaning of this
term, so it does not \emph{classify} particular objects. Therefore, it does not provide
information which particular object should be included to a family and which is not (in
particular, whether a particular object belongs to a particular cluster or is from the
background). The purpuse of the wavelet analysis is to analyse the statistical distribution
as a smooth function and to detect unusual patterns inside it.
\end{enumerate}

Therefore, we may argue that the wavelet analysis was undeservedly abandoned in this task
over years. It can be used as an independent method of cluster detection, in particular in
the asteroid families search, but it also needs further development.

\begin{acknowledgements}
RVB acknowledges the support of Russian Science Foundation grant 18-12-00050 for the
programming and asteroid data analysis work. EIR was supported by Russian Foundation for
Basic Research grant 17-02-00542~A for 3D interpretation of asteroid groups and for their
cross-comparison with known asteroid families. The authors would like to express gratitude
to the reviewers of the manuscript, Prof. Valerio Carruba and an anonymous one, for their
useful comments and suggestions.
\end{acknowledgements}


\bibliographystyle{spbasic}      % basic style, author-year citations
\bibliography{wavaster}   % name your BibTeX data base

\begin{thebibliography}{20}
\providecommand{\natexlab}[1]{#1}
\providecommand{\url}[1]{{#1}}
\providecommand{\urlprefix}{URL }
\expandafter\ifx\csname urlstyle\endcsname\relax
  \providecommand{\doi}[1]{DOI~\discretionary{}{}{}#1}\else
  \providecommand{\doi}{DOI~\discretionary{}{}{}\begingroup
  \urlstyle{rm}\Url}\fi
\providecommand{\eprint}[2][]{\url{#2}}

\bibitem[{Baluev(2018)}]{Baluev18a}
Baluev R.~V.: Statistical detection of patterns in unidimensional distributions
  by continuous wavelet transforms. \ac 23, 151--165 (2018)

\bibitem[{Baluev and Shaidulin(2018)}]{Baluev18b}
Baluev R.~V., Shaidulin V.~S.: Fine-resolution wavelet analysis of exoplanetary
  distributions: hints of an overshooting iceline accumulation. \apss 363, 192
  (2018)

\bibitem[{Baluev et~al(2020{\natexlab{a}})Baluev, Rodionov, and
  Shaidulin}]{BalRodShai19}
Baluev R.~V., Rodionov E.~I., Shaidulin V.~S.: Isotropic wavelet denoising
  algorithm for bivariate density analysis and estimation. preprint
  arXiv.org:1903.10167 (2020{\natexlab{a}})

\bibitem[{Baluev et~al(2020{\natexlab{b}})Baluev, Shaidulin, and
  Veselova}]{BalShaiVes20}
Baluev R.~V., Shaidulin V.~S., Veselova A.~V.: High-velocity moving groups in
  the {S}olar neighborhood in {GAIA DR2}. Acta Astron. (accepted)
  (2020{\natexlab{b}})

\bibitem[{Bro{\v{z}} and Vokrouhlick{\'y}(2008)}]{Broz08}
Bro{\v{z}} M., Vokrouhlick{\'y} D.: Asteroid families in the first-order
  resonances with {J}upiter. \mnras 390, 715--732 (2008)

\bibitem[{Carruba et~al(2019)Carruba, Aljbaae, and Lucchini}]{Carruba19}
Carruba V., Aljbaae S., Lucchini A.: Machine-learning identification of
  asteroid groups. \mnras 488, 1377--1386 (2019)

\bibitem[{Hirayama(1918)}]{Hirayama18}
Hirayama K.: Groups of asteroids probably of common origin. \aj 31, 185--188
  (1918)

\bibitem[{Hirayama(1922)}]{Hirayama22}
Hirayama K.: Families of asteroids. Japanese J. Astron. \& Geophys. 1, 55--93
  (1922)

\bibitem[{Kne{\v{z}}evi{\'c} and Milani(2003)}]{Knezevic03}
Kne{\v{z}}evi{\'c} Z., Milani A.: Proper element catalogs and asteroid
  families. \aap 403, 1165--1173 (2003)

\bibitem[{Kne{\v{z}}evi{\'c} et~al(2002)Kne{\v{z}}evi{\'c}, Lema{\^\i}tre, and
  Milani}]{Knezevic02}
Kne{\v{z}}evi{\'c} Z., Lema{\^\i}tre A., Milani A.: (2002) The determination of
  asteroid proper elements. In: Bottke W.~F., Cellino A., Paolicchi P., Binzel
  R.~P. (eds) Asteroids III, University of Arizona Press, Tucson, chap 5.1, pp
  603--612

\bibitem[{Masiero et~al(2015)Masiero, {D}e{M}eo, Kasuga, and
  H.~Parker}]{Masiero15}
Masiero J., {D}e{M}eo F., Kasuga T., H.~Parker A.~H.: (2015) Asteroid family
  physical properties. In:  \cite{AsteroidsIV}, chap 2.3, pp 323--340

\bibitem[{Michel et~al(2015)Michel, DeMeo, and Bottke}]{AsteroidsIV}
Michel P., DeMeo F.~E., Bottke W.~F. (eds)  Asteroids IV. University of Arizona
  Press, Tucson (2015)

\bibitem[{Milani et~al(2014)Milani, Cellino, Kne\v{z}evi\'{c}, Novakovi\'{c},
  Spoto, and Paolicchi}]{Milani14}
Milani A., Cellino A., Kne\v{z}evi\'{c} Z., Novakovi\'{c} B., Spoto F.,
  Paolicchi P.: Asteroid families classification: {E}xploiting very large
  datasets. Icarus 239, 46--73 (2014)

\bibitem[{Milani et~al(2017)Milani, Kne{\v{z}}evi{\'c}, Spoto, Cellino,
  Novakovi{\'c}, and Tsirvoulis}]{Milani17}
Milani A., Kne{\v{z}}evi{\'c} Z., Spoto F., Cellino A., Novakovi{\'c} B.,
  Tsirvoulis G.: On the ages of resonant, eroded and fossil asteroid families.
  Icarus 288, 240--264 (2017)

\bibitem[{Murray and Dermott(1999)}]{MurrayDermott}
Murray C.~D., Dermott S.~F.: Solar System Dynamics. Cambridge University Press
  (1999)

\bibitem[{Nesvorn{\'y} et~al(2002)Nesvorn{\'y}, Jr, Dones, and
  Levison}]{Nesvorny02}
Nesvorn{\'y} D., Jr W.~F.~B., Dones L., Levison H.~F.: The recent breakup of an
  asteroid in the main-belt region. Nature 417, 720--722 (2002)

\bibitem[{Nesvorn{\'y} et~al(2015)Nesvorn{\'y}, Bro{\v{z}}, and
  Carruba}]{Nesvorny15}
Nesvorn{\'y} D., Bro{\v{z}} M., Carruba V.: (2015) Identification and dynamical
  properties of asteroid families. In:  \cite{AsteroidsIV}, chap 2.3, pp
  297--321

\bibitem[{Snodgrass et~al(2012)Snodgrass, Carry, Dumas, and
  Hainaut}]{Snogdrass12}
Snodgrass C., Carry B., Dumas C., Hainaut O.: Characterisation of candidate
  members of (136108) {H}aumea's family. \aap 511, A72 (2012)

\bibitem[{Zappal{\`a} et~al(1990)Zappal{\`a}, Cellino, Farinella, and
  Kne{\v{z}}evi{\'c}}]{Zappala90}
Zappal{\`a} V., Cellino A., Farinella P., Kne{\v{z}}evi{\'c} Z.: Asteroid
  families. {I}. {I}dentification by hierarchical clustering and reliability
  assessment. \aj 100, 2030--2046 (1990)

\bibitem[{Zappal{\`a} et~al(1995)Zappal{\`a}, Bendjoya, Cellino, Farinella, and
  Froeschl{\'e}}]{Zappala95}
Zappal{\`a} V., Bendjoya P., Cellino A., Farinella P., Froeschl{\'e} C.:
  Asteroid families: {S}earch of a 12487 asteroid sample using two different
  clustering techniques. Icarus 116, 291--314 (1995)

\end{thebibliography}

\end{document}